\documentclass[aps,prl,twocolumn,superscriptaddress]{revtex4-2}
\usepackage{graphicx}
\usepackage{amsmath}
\usepackage{amssymb}
\usepackage{mathtools}
\usepackage{xcolor}
\usepackage{braket}
\usepackage{soul}
\usepackage[colorlinks=true, citecolor=blue, urlcolor=blue, linkcolor=blue,bookmarks=false,hypertexnames=true]{hyperref} 
\usepackage{appendix}
\usepackage{array}
\usepackage[column=O]{cellspace}
\newcolumntype{P}[1]{>{\centering\arraybackslash}p{#1}}
\allowdisplaybreaks
\usepackage{lipsum}
\usepackage{acro}
\usepackage{enumitem}
\usepackage{amstext}

\DeclareAcronym{TMDC}{short=TMDC, long=transition metal dichalcogenide,}
\DeclareAcronym{DFT}{short=DFT, long=density functional theory,}
\DeclareAcronym{mDF}{short=mDF, long=massive Dirac Fermion,}
\DeclareAcronym{2D}{short=2D, long=two-dimensional,}
\DeclareAcronym{3D}{short=3D, long=three-dimensional,}
\DeclareAcronym{VB}{short=VB, long=valence band,}
\DeclareAcronym{CB}{short=CB, long=conduction band,}
\DeclareAcronym{UC}{short=UC, long=unit cell,}
\DeclareAcronym{PBE}{short=PBE, long=Perdew-Burke-Ernzerhof,}
\DeclareAcronym{PAW}{short=PAW, long=projector augmented wave,}
\DeclareAcronym{GGA}{short=GGA, long=generalized gradient approximation,}
\DeclareAcronym{SOC}{short=SOC, long=spin-orbit coupling,}
\DeclareAcronym{BSE}{short=BSE, long=Bethe-Salpeter equation,}
\DeclareAcronym{hBN}{short=hBN, long=hexagonal boron nitride,}
\DeclareAcronym{BZ}{short=BZ, long=Brillouin zone,}
\DeclareAcronym{NN}{short=NN, long=nearest neighbour,}
\DeclareAcronym{NNN}{short=NNN, long=next-nearest neighbour,}
\DeclareAcronym{LL}{short=LL, long=Landau level,}
\DeclareAcronym{XC}{short=XC, long=exchange correlation,}
\DeclareAcronym{TDSE}{short=TDSE, long=time-dependent Schr\"odinger equation,}
\DeclareAcronym{X0}{short=$X^0$, long=neutral exciton,}
\DeclareAcronym{X-}{short=$X^-$, long=negatively charged exciton,}
\DeclareAcronym{X*}{short=$X^*$, long=neutral exciton interacting with the filled CB,}
\DeclareAcronym{GS}{short=GS, long=ground state,}
\DeclareAcronym{DME}{short=DME, long=dipole matrix element,}
\DeclareAcronym{HO}{short=HO, long=harmonic oscillator}
\DeclareAcronym{2DEG}{short=2DEG, long=\ac{2D} electron gas}
\DeclareAcronym{CI}{short=CI, long=configuration interaction}
\DeclareAcronym{SM}{short=SM, long=Supplementary Material}

\begin{document}
\preprint{APS/123-QED}

\title{Magnetoexcitons and Massive Dirac Fermions in  Monolayers of Transition Metal Dichalcogenides in a High Magnetic Field}
\date{\today}

\author{Katarzyna Sadecka}
%\email{sadkat27@gmail.com}
\affiliation{Department of Physics, University of Ottawa, Ottawa, K1N6N5, Canada}
\affiliation{Institute of Theoretical Physics, Wroc\l aw University of Science and Technology, Wybrze\.ze Wyspia\'nskiego 27, 50-370 Wroc\l aw, Poland}

\author{Marek Korkusiński}
\affiliation{Department of Physics, University of Ottawa, Ottawa, K1N6N5, Canada}
\affiliation{Quantum and Nanotechnologies Research Centre, National Research Council, Ottawa K1A0R6, Canada}

\author{Ludmi\l{}a Szulakowska}
\affiliation{Department of Physics, University of Ottawa, Ottawa, K1N6N5, Canada}

\author{Pawe\l{} Hawrylak}
\affiliation{Department of Physics, University of Ottawa, Ottawa, K1N6N5, Canada}

%%%%%%%%%%%%%%%%%%%%%%%%%%%%%%%%%%%%%%%%% Abstract %%%%%%%%%%%%%%%%%%%%%%%%%%%%%%%%%%%%%%%%%
\begin{abstract} 
We present a theory of the emission spectrum of magnetoexcitons interacting with a $\nu = 1$ quantum Hall state of massive Dirac fermions in monolayer transition metal dichalcogenides in high magnetic fields. Using an \textit{ab initio}--parametrized massive Dirac fermion model including valley and spin degrees of freedom, combined with exact diagonalization techniques, we show that interband emission from the massive Dirac Fermion magnetoexciton interacting with $\nu = 1$ state directly probes intra--conduction-band excitations of the $\nu = 1$. Many-body interactions with the filled massive Dirac fermion $\nu = 1$ level yield a strong renormalization of the emission spectrum, including fully polarized emission, a pronounced redshift, and broadening relative to neutral and charged excitons. The calculated spectra are consistent with recent experiments~\cite{Finley2021MoS2Magnetic, Finley2022MoS2Magnetic, Oreszczuk_2023}, establishing magneto-spectroscopy as a probe of finite carrier densities in massive Dirac systems.
\end{abstract}
\maketitle
%%%%%%%%%%%%%%%%%%%%%%%%%%%%%%%%%%%%%%%%%%%%%%%%%%%%%%%%%%%%%%%%%%%%%%%%%%%%%%%%%%%%%%%%%%

%%%%%%%%%%%%%%%%%%%%%%%%%%%%%%%%%%%%%%%%%%%%%%%%%%%%%%%%%%%%

%%%%%%%%%%%%%%%%%%%%%%%%%%%%%%%%%%%%%%
\textit{Introduction}---
Understanding the optical properties of atomically thin semiconductors, particularly \acp{TMDC}, with controlled carrier densities enables detection of free carriers and their density. Early studies of quasi-\ac{2D} semiconductor layers showed that photon absorption creates excitons, whose interactions with free carriers can be described in terms of Fermi edge singularities, the Anderson orthogonality catastrophe~\cite{Skolnick1987, Hawrylak1991FermiEdge, Brown1996, Brum1996}, and the Fermi polaron model~\cite{Sidler2017}. Subsequent work extended these studies to high magnetic fields~\cite{Finkelstein1997, Hawrylak19972DEG, Gravier1998, BarJoseph2005, Byszewski2006, Plochocka2007, Nomura2014}, tracking the evolution of \ac{2DEG} emission spectra with carrier density. Effects of incompressible Laughlin liquids in fractionally filled \acp{LL} were interpreted as trions formed by excitons bound to fractionally charged quasiparticles~\cite{Byszewski2006}, while recombination of optically generated holes with electrons in partially filled \acp{LL} produced discontinuities near integer filling factors in GaAs-based \ac{2DEG} systems~\cite{Potemski1991, Hawrylak19972DEG, Gravier1998, Byszewski2006, Nomura2014, trushin2016optical}. These features, attributed to Anderson-Fano-like resonances and interaction-driven spectral weight redistribution~\cite{Hawrylak19972DEG}, established optical spectroscopy as a sensitive probe of electronic correlations in quantum Hall systems.

Recent studies have applied these techniques to atomically thin \acp{TMDC}~\cite{Sidler2017,Finley2021MoS2Magnetic,Finley2022MoS2Magnetic,Oreszczuk_2023}, examining the evolution of recombination spectra with carrier density in high magnetic fields. Spectra show a progression from magnetoexciton emission to magnetotrion features, followed by a broad low-energy signal. These observations were interpreted using an \textit{ab initio} exciton-trion model developed at zero field and extrapolated to finite fields via Zeeman shifts renormalized by exchange interactions with a resident electron gas, assumed localized by disorder and described through Brillouin-function expressions analogous to magnetic semiconductors.

%%%%%%%%%%%%%%%%% Fig %%%%%%%%%%%%%%%%
\begin{figure}[b]
\includegraphics[width=\linewidth]{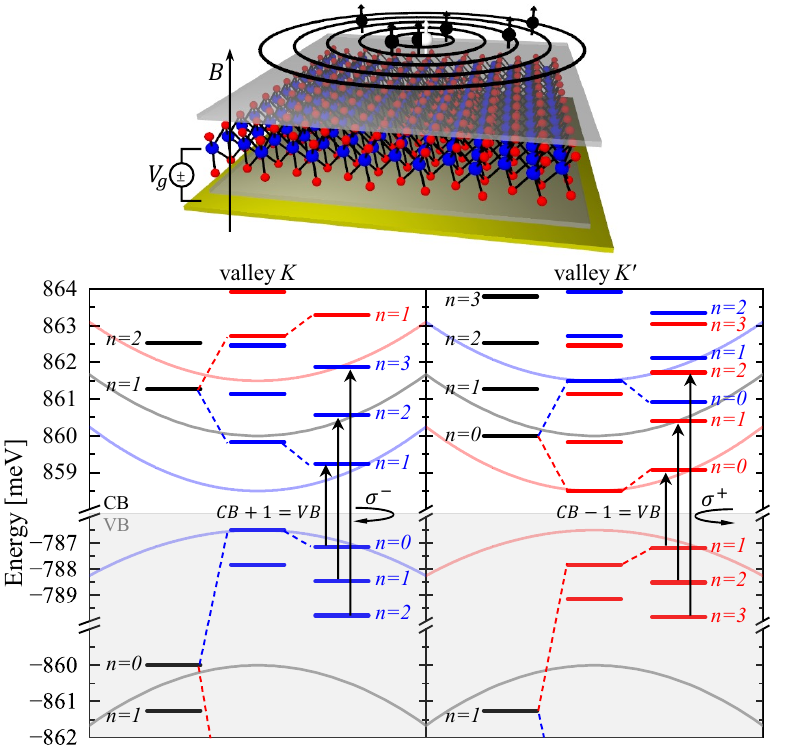}
    \caption{Top: monolayer MoS$_2$ in a magnetic field, illustrating cyclotron motion.  
    Bottom: \ac{LL} spectra at $B=10$~T for $K$ and $K'$ valleys, showing three model stages: (i) no \ac{SOC}, (ii) \ac{SOC} only, (iii) full spin–valley coupling ($g_e=-2$, $g_h=-2.2$). Background lines: $B=0$ bands (grey: no \ac{SOC}, red/blue: spin-split). Dashed lines: selected \acp{LL} labeled by $n$. Vertical arrows ($\sigma^\pm$) indicate optically allowed interband transitions.}
\label{fig:TMDCwithLL}
\end{figure}
%%%%%%%%%%%%%%%%%%%%%%%%%%%%%%%%%%%%%%

In this work, we elucidate the microscopic origin of experimentally observed spectra~\cite{Finley2021MoS2Magnetic, Finley2022MoS2Magnetic, Oreszczuk_2023} by accounting for Landau-level formation in a disorder-free system~\cite{TuanDery2025}. Starting from \textit{ab initio} calculations for MoS$_2$, we extract an effective mass model (\ac{mDF}) whose magnetic field dependence follows Goerbig \textit{et al.}~\cite{rose2013spin} and has been experimentally confirmed~\cite{Plochocka2015, Koperski2019, Stier2016, Scrace2015}. The \ac{mDF} model captures two non-equivalent valleys with spin-locked bands and features a degenerate $n=0$ conduction \ac{LL} in valley $K'$ and $n=0$ valence \ac{LL} in valley $K$. Using exact diagonalization, we predict the spectra of excitons interacting with \acp{mDF}, focusing on a neutral exciton coupled to the fully filled lowest \ac{LL} ($\nu=1$). This regime directly probes the intra-conduction-band excitations of the $\nu=1$ state, with the neutral exciton and negatively charged trion serving as reference emission spectra.

%%%%%%%%%%%%%%%%%%%%%%%%%%%%%%%%%%%%%%%%%%%%%%%%%%%%%%%%%%%%

%%%%%%%%%%%%%%%%%%%%%%%%%%%%%%%%%%%%%%
\vspace{6pt}\textit{Non-interacting massive Dirac fermions}---
Using \textit{ab initio} calculations the electronic properties of monolayer MoS$_2$ in the 2H phase were determined, yielding a direct $K$--$K$ band gap $\Delta = 1.27$~eV, spin-orbit splittings $\Delta^{\rm CB}_{\rm SOC}=-3$~meV and $\Delta^{\rm VB}_{\rm SOC}=147$~meV, and strong spin--valley locking, particularly in the \ac{VB}. The lowest-energy conduction and highest-energy valence subbands in each valley share the same spin, producing bright, spin-allowed transitions.

These low-energy bands are captured by the \ac{mDF} model~\cite{bradlyn2016beyond,Szulakowska2019magnetoX,bieniek2022nanomaterials,Sadecka2025heteroQD}, parametrized by the gap, Fermi velocity, and \ac{SOC} splittings. This model provides a tractable platform for describing optical properties of \acp{TMDC} in magnetic fields and varying carrier densities~\cite{xiao2012coupled,Kormanyos2013,VanDerDonck2018}.

A perpendicular magnetic field $B$ is included via the Peierls substitution $\vec{p} \to \vec{p}+|e|\vec{A}$ (symmetric gauge). The sample is modeled as a monolayer encapsulated in \ac{hBN} and electrostatically doped via a metallic gate, as in recent experiments (top panel in Fig.~\ref{fig:TMDCwithLL}). This setup yields a block-diagonal Hamiltonian in the spin-resolved basis $\{\text{CB}\downarrow,\text{VB}\downarrow,\text{CB}\uparrow,\text{VB}\uparrow\}$, which for both valleys can be written as:
%%%%%%%%%%%%%%%%%%%%%%%%%%%%%
\begin{equation}
\begin{split}
    &\hat{H}_B^{(s)}(K) =
    \begin{bmatrix}
    \frac{\Delta + s\Delta_{\rm SOC}^{\rm CB} - sg_e \mu_B B}{2} & -i c \hat{a} \\
    i c \hat{a}^\dagger & \frac{-\Delta - s\Delta_{\rm SOC}^{\rm CB} - sg_e \mu_B B}{2}
    \end{bmatrix}, \\
    &\hat{H}_B^{(s)}(K') =
    \begin{bmatrix}
    \frac{\Delta + s\Delta_{\rm SOC}^{\rm CB} - sg_e \mu_B B}{2} & -i c \hat{a}^\dagger \\
    i c \hat{a} & \frac{-\Delta - s\Delta_{\rm SOC}^{\rm CB} - sg_e \mu_B B}{2}
    \end{bmatrix},
\end{split}
\label{eq:mDFmagneticH}
\end{equation}
%%%%%%%%%%%%%%%%%%%%%%%%%%%%%
with $s=\pm1$ denoting spin $\downarrow/\uparrow$, $c=v_F\sqrt{2}/l_B$, $\mu_B$ the Bohr magneton, $l_B = \sqrt{\hbar/eB}$ the magnetic length, and the Lande factors $g_e=-2$, $g_h=-2.2$~\cite{Stier2021PRR}. The Zeeman effect shifts the diagonal terms linearly with $B$, while the off-diagonal orbital terms produce discrete \acp{LL}. The resulting eigenenergies and spinor wavefunctions can be expressed as
%%%%%%%%%%%% Eq %%%%%%%%%%%%%
    $\ket{\psi^{\rm VB/CB}_{n,m}(K)} = \alpha^{\rm VB/CB}_n \ket{n-1,m} + \beta^{\rm VB/CB}_n \ket{n,m}$ 
    and 
    $\ket{\psi^{\rm VB/CB}_{n,m}(K')} = \alpha^{\rm VB/CB}_n \ket{n,m} + \beta^{\rm VB/CB}_n \ket{n-1,m}$,
%%%%%%%%%%%%%%%%%%%%%%%%%%%%%
where $\ket{n,m}$ are the \ac{2D} harmonic oscillator states and $m$ labels the orbital degeneracy, producing macroscopic \ac{LL} degeneracy. Details of derivation of the single-particle \ac{mDF} Hamiltonians are given in the \ac{SM}.

The \ac{LL} spectrum at $B = 10$~T is shown in Fig.~\ref{fig:TMDCwithLL} for both $K$ and $K'$ valleys. Three successive ladders illustrate the effects of increasing complexity: (i) without \ac{SOC} (black), showing basic Dirac quantization and valley Zeeman splitting, (ii) including \ac{SOC} (light red/blue), producing spin–valley–locked \acp{LL}, and (iii) including finite $g$-factors, giving a fully spin- and valley-resolved spectrum. Notably, the $n=0$ \ac{CB} \ac{LL} exists only in $K'$ and the $n=0$ \ac{VB} only in $K$, deriving solely from the diagonal Hamiltonian terms and insensitive to the off-diagonal diamagnetic coupling.

%%%%%%%%%%%%%%%%%%%%%%%%%%%%%%%%%%%%%%%%%%%%%%%%%%%%%%%%%%%%

%%%%%%%%%%%%%%%%%%%%%%%%%%%%%%%%%%%%%%
\vspace{6pt}\textit{Interband magnetoexcitons in the presence of $\nu=1$}---
The single-particle \ac{LL} orbitals are populated with a total of $N = N_V + N_G$ electrons, where $N_V$ corresponds to the electrons filling the \ac{VB} \acp{LL}, and $N_G$ denotes the additional electrons brought into the \ac{CB} by the gate potential. The Hamiltonian of the interacting electrons can be written as
%%%%%%%%%%%% Eq  Hamiltonian  %%%%%%%%%%%%%
\begin{equation}
\begin{split}
    \hat{H} =& \sum_i E_i
    \hat{c}_i^\dagger\hat{c}_i
    + \frac{1}{2}\sum_{i,j,k,l} \braket{i,j|V_C|k,l} \hat{c}_i^\dagger\hat{c}_j^\dagger\hat{c}_k\hat{c}_l 
     \\
    &\quad -\sum_{i,j} \braket{i|V_P|j}
    \hat{c}_i^\dagger\hat{c}_j.
\end{split}
\label{eq:InteractingHamiltonian}
\end{equation}
%%%%%%%%%%%%%%%%%%%%%%%%%%%%%
Here, the composite index $i = \{n, m, \tau, s, \text{CB/VB}\}$ encompasses the \ac{LL} orbital indices, valley, spin, and band index of the single-particle state (similarly for $j, k, l$). $E_i$ denotes the single-particle energy of the \ac{LL} orbital $i$. The Coulomb matrix elements $\braket{i,j|V_C|k,l}$ describe the  interaction between electrons. Details of their calculation are provided in the \ac{SM}. The scattering matrix elements $\braket{i|V_P|j}$ account for the interaction between electrons and the positive background, which ensures the overall charge neutrality of the system.

%%%%%%%%%% Fig configs %%%%%%%
\begin{figure}[t]
\centering\includegraphics[width=\linewidth]{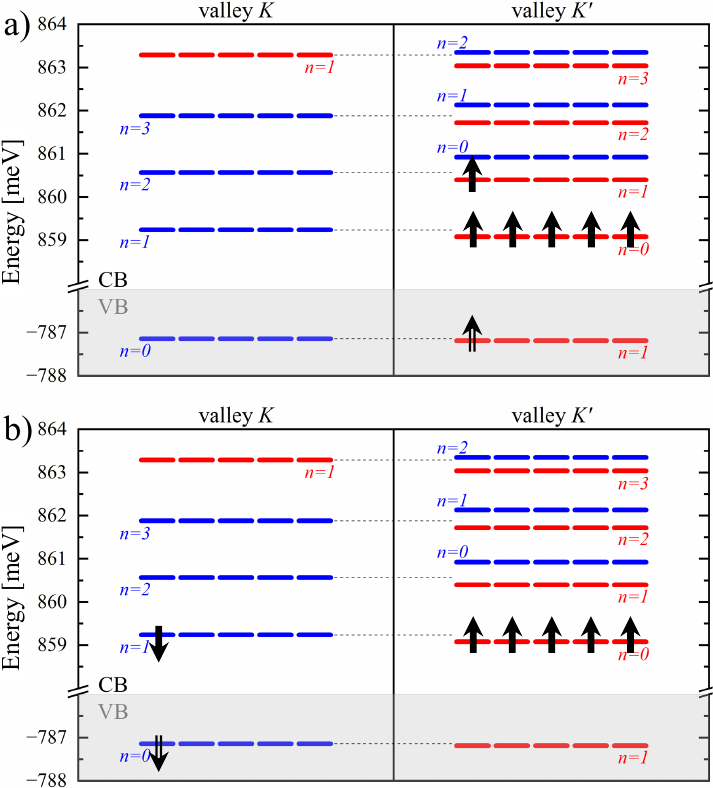}
    \caption{Schematic of a neutral exciton (a) in valley $K'$ and (b) in valley $K$ interacting with the $\nu=1$ state at $B=10$~T. Black arrows in $n=0$ \ac{LL} denote the fully filled $\nu=1$ state; a model degeneracy of 5 is used.}
\label{fig:2DEG_SpinConfs}
\end{figure}
%%%%%%%%%%%%%%%%%%%%%%%%%%%%%

Building on earlier work on optical detection of integer~\cite{Hawrylak19972DEG} and fractional quantum Hall states~\cite{Byszewski2006,Nomura2014}, we focus on the interband magnetoexciton interacting with the fully filled lowest \ac{LL}, i.e., the $\nu=1$ state. The system contains $N = N_V + N_G$ electrons, where $N_G$ equals the \ac{LL} degeneracy $M_{LL}$. The eigenstates and eigenenergies of the magnetoexciton are obtained using a \ac{CI} approach in a basis of electron–hole pair excitations built on the reference \ac{GS}, in which one electron is promoted from the \ac{VB} to the \ac{CB}. After recombination, the system is left in a final state with a filled \ac{VB} and intraband pair excitations in the \ac{CB}. The final states (intraband excitons) are obtained by diagonalizing the Hamiltonian in this restricted basis, and the emission spectra are constructed using Fermi’s golden rule. Detailed information on the \ac{CI} approach is provided in the \ac{SM}.

We start from the initial emission state in a magnetic field $B=10$~T, with the reference configuration $\ket{GS}$ consisting of a fully occupied $n=0$ \ac{LL} in $K'$ and a filled \ac{VB}. All excitation energies are measured relative to its total energy (including interactions). Electron–hole pair excitations basis is then generated as $\ket{i,j} = \hat{c}^\dagger_i \hat{c}_j \ket{GS}$, with interband excitations arising when $i$ ($j$) belongs to the \ac{CB} (\ac{VB}). The lowest-energy families are formed by considering the top \ac{VB} and lowest \ac{CB} \acp{LL} in each valley.

In Fig.~\ref{fig:2DEG_SpinConfs}, the optically active pair configurations are shown schematically. Spin-up (down) \acp{LL} are represented by red (blue) horizontal bars, with a model degeneracy $M_{LL}=5$ used for illustration ($M_{LL}=31$ in calculations). Single (double) arrows mark electrons (holes) in the \ac{CB} (\ac{VB}). Panel (a) shows a $K'$-valley excitation with a hole in $n=1$ \ac{VB} \ac{LL} and an electron in $n=1$ \ac{CB} \ac{LL}, yielding $M_{LL}^2$ configurations. Panel (b) shows a $K$-valley excitation with a hole in $n=0$ \ac{VB} and an electron in $n=1$ \ac{CB} \ac{LL} (spins opposite to panel a), also giving $M_{LL}^2$ configurations.

Here, many-particle configurations are restricted to a single \ac{LL}. Including additional \acp{LL} increases the Hilbert-space size factorially, but previous studies~\cite{TuanDery2025} show that this only produces quantitative corrections without changing the qualitative behavior. Our choice thus captures the essential physics while remaining computationally tractable.

\begin{figure}[t]
\centering\includegraphics[width=\linewidth]{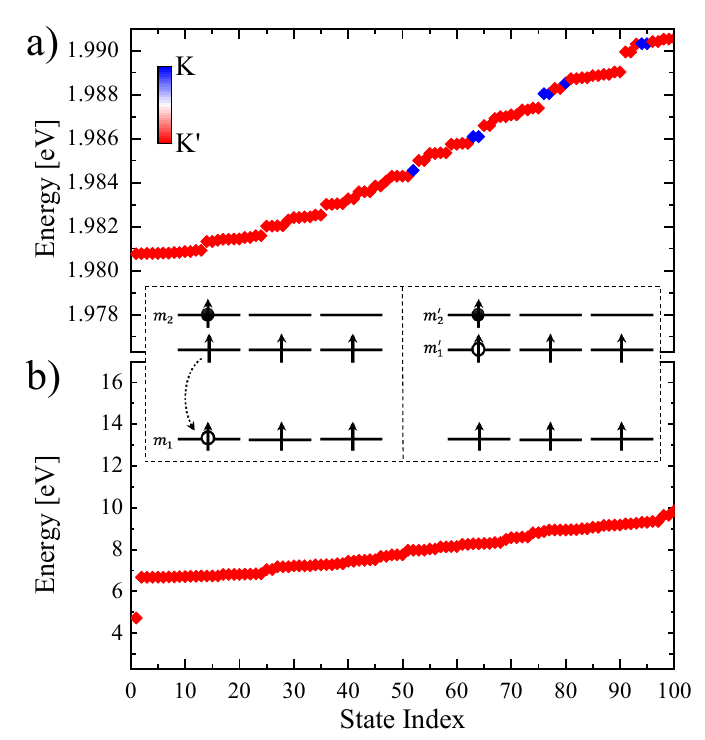}
    \caption{Excitonic spectra of (a) an interband exciton interacting with $\nu=1$ and (b) an intraband exciton (final emission state), displayed on a common energy scale; colours denote valley. The inset sketches the recombination process from the interband exciton to the intraband exciton.}
\label{fig:Xcomparison}
\end{figure}

The spectra of correlated exciton states, $\ket{\Psi_k^{(\mathrm{ini})}}=\sum_{i,j} A_{i,j}^{(k)}\ket{i,j}$, are characterized by configuration mixing, which leads to a broadening of the energy manifolds in each valley and results in an overlap between the $K$- and $K'$-valley exciton families. Despite this overlap, the \ac{GS} and several lowest-energy excited states remain primarily composed of $K'$-valley configurations. This energetic ordering originates from exchange interactions between the additional \ac{CB} electron forming the exciton and the fully occupied $\nu=1$ \ac{LL}. For a $K'$-valley exciton, the electron shares the same valley and spin quantum numbers as the electrons in the $\nu=1$ state, allowing for a finite exchange contribution that lowers its energy. In contrast, such exchange interaction is absent for a $K$-valley exciton, where the \ac{CB} electron resides in the opposite valley. As a result, $K'$-valley excitons are energetically favored. Figure~\ref{fig:Xcomparison}(a) shows the energies of the lowest $100$ excitonic states as a function of the state index. Among the first $50$ states, only $K'$-valley excitons (red) are present, while the lowest-energy $K$-valley exciton (blue) appears as the $51$st state.

%%%%%%%%%%%%% Fig %%%%%%%%%%%
%%%%%%%%%%%%%%%%%%%%%%%%%%%%%

%%%%%%%%%%%%%%%%%%%%%%%%%%%%%%%%%%%%%%
%\vspace{6pt}\textit{Emission spectra}---
We now consider the radiative recombination of a valence hole with a \ac{CB} electron. The final states after emission are identified according to the optical selection rules (see \ac{SM}): a transition is allowed for an electron–hole pair $\ket{i,j}$ if both carriers occupy the same valley, have identical spin, differ in \ac{LL} index $n$ by 1, and share the same orbital index $m$. These rules are applied to interband exciton states in both valleys.

For the $K$-valley magnetoexciton [Fig.~\ref{fig:2DEG_SpinConfs}(b)], the hole recombines only with the electron in the same valley (both in blue), while the remaining electrons stay in $\ket{GS}$. No additional pair excitations at this energy arise, so $\ket{GS}$ is the final eigenstate.

In contrast, recombination in the $K'$ valley [Fig.~\ref{fig:2DEG_SpinConfs}(a)] is more complex. The optical rules require the hole to recombine with an electron in the $n=0$ \ac{LL} forming the $\nu=1$ state (dotted arrow), leaving a vacancy in the filled \ac{CB} \ac{LL}, as shown in the inset of Fig.~\ref{fig:Xcomparison}. This necessitates intraband pair configurations $\ket{i,j}$ within the \ac{CB}, with the electron on orbital $i$ of the $n=1$ \ac{LL} and the hole on orbital $j$ of the $n=0$ \ac{LL}; there are $M_{LL}^2$ such configurations. Diagonalization yields intraband exciton states $\ket{\Psi_q^{(fin)}} = \sum_{i,j} B^{(q)}_{i,j} \ket{ij}$, which serve as the final states in the $K'$ emission spectra. Figure~\ref{fig:Xcomparison}(b) shows the energies of the first $100$ intraband excitons in the same window as the interband excitons [Fig.~\ref{fig:Xcomparison}(a)]. These intraband energies are lower by roughly the bandgap, and their dispersion is notably flat, with the first $100$ states lying within $6$~meV, compared to $10$~meV for the interband excitons.

%%%%%%%%%%%%% Fig %%%%%%%%%%%
\begin{figure}[t!]
\centering\includegraphics[width=\linewidth]{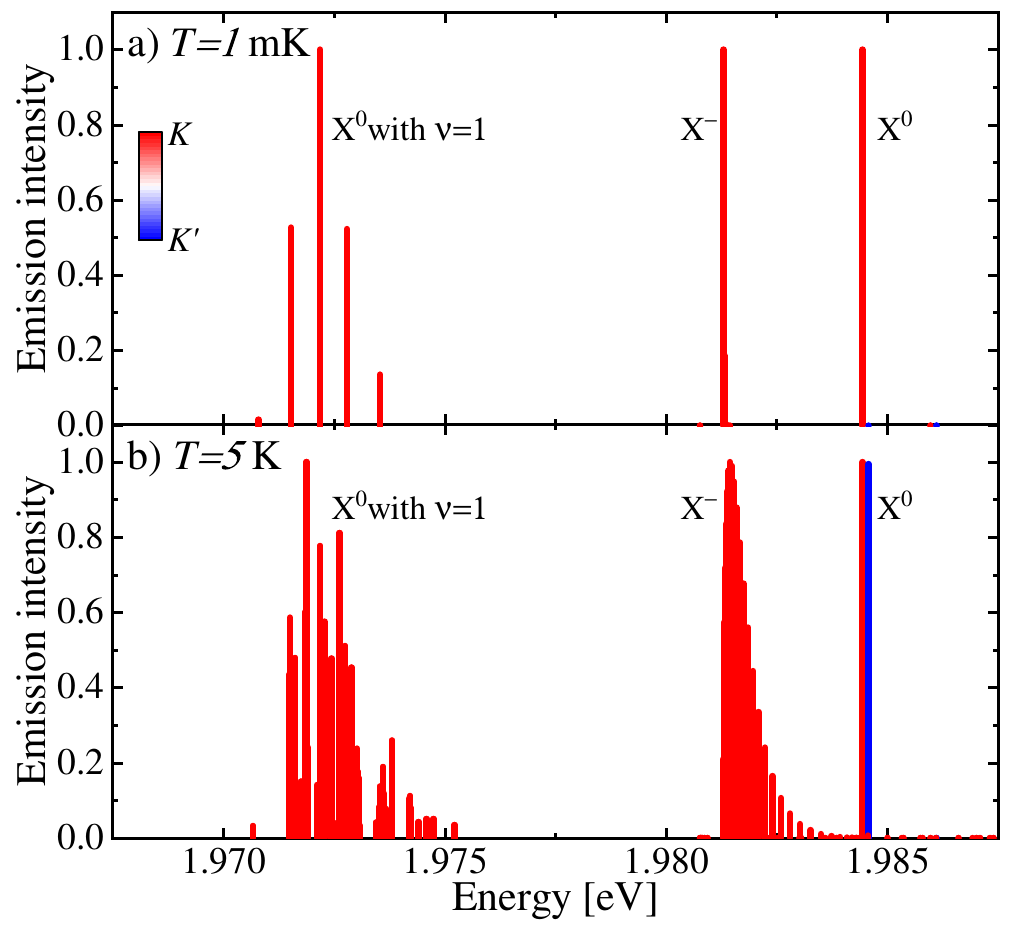}
    \caption{Emission spectra of $X^0$, $X^-$ (see \ac{SM}), and an exciton interacting with $\nu=1$ in a MoS$_2$ monolayer at $B=10$~T, shown for (a) $T=1$~mK and (b) $T=5$~K.}
\label{fig:Emission}
\end{figure}
%%%%%%%%%%%%%%%%%%%%%%%%%%%%%

Having obtained the initial interband exciton states $\ket{\Psi^{(ini)}_k}$ and the final intraband states $\ket{\Psi^{(fin)}_q}$, the emission spectrum is calculated using Fermi's Golden Rule 
$A(\hbar\omega) = \sum_{k,q} W_k \left| \braket{\Psi^{(fin)}_q | \hat{P} | \Psi^{(ini)}_k} \right|^2 \delta(E_k^{(ini)} - E_q^{(fin)} - \hbar\omega)$,
where $W_k$ defines thermal occupations at temperature $T$. The spectrum is thus a weighted sum of peaks at photon energies $\hbar\omega = E_k^{(ini)} - E_q^{(fin)}$, with amplitudes set by the matrix elements of the interband polarization operator 
$\hat{P} = \sum_{i,j} D_{i,j} \hat{c}^\dagger_j \hat{c}_i$, 
which remove an electron from single-particle state $i$ in the \ac{CB} and place it in state $j$ in the \ac{VB}. The transition matrix elements $D_{i,j}$ encode the optical selection rules and determine the oscillator strength. Their absolute normalization factors $D_{1,0}$ and $D_{0,1}$ are obtained from the single-particle model. The details of the calculations are provided in the \ac{SM}. The central result of this work is to show how the presence of the $\nu=1$ state and associated correlations manifest in $A(\hbar\omega)$.

For the $K$ valley [Fig.~\ref{fig:2DEG_SpinConfs}(b)], only a single final state $\ket{GS}$ exists. The dipole element is $D_{i,j} = D_{1,0}\,\delta(m_i,m_j)$, nonzero only if the electron in $n=1$ \ac{LL} and the hole in $n=0$ \ac{LL} share the same orbital $m$. The corresponding matrix element is 
$\braket{GS | \hat{P} | \Psi^{(ini)}_k} = D_{1,0} \sum_{i,j} A_{i,j}^{(k)} \delta(m_i,m_j)$, 
where among the $M_{LL}^2$ pair configurations, only the $M_{LL}$ “diagonal” contributions with aligned electron and hole yield a nonzero term.

In the $K'$ valley, both initial and final states belong to a manifold of many-body excitations, so the emission reflects transitions into a quasi-continuum and correlation effects appear. The dipole element is similarly defined: $D_{i,j} = D_{0,1}\,\delta(m_i,m_j)$, with the hole in $n=1$ \ac{LL} and the electron in $n=0$ \ac{LL}. All intervalley pair configurations are optically active, since each hole finds an electron in the filled zeroth \ac{LL} to recombine with. The corresponding matrix element is
%%%%%%%%%%%%%%%%%%%%%%%%%%%%%
\begin{equation}
\begin{split}
   &\braket{\Psi^{(fin)}_q | \hat{P} | \Psi^{(ini)}_k} = \\
    &\quad D_{0,1} \sum_{i, j_\mathrm{CB}, j_\mathrm{VB}} \left(B_{i, j_\mathrm{CB}}^{(q)}\right)^* A_{i, j_\mathrm{VB}}^{(k)} \delta(m_{j_\mathrm{CB}}, m_{j_\mathrm{VB}}),
\end{split}
\end{equation}
%%%%%%%%%%%%%%%%%%%%%%%%%%%%%
where $j_\mathrm{VB}$ tracks the hole orbital in the \ac{VB} for the initial state and $j_\mathrm{CB}$ tracks the hole orbital in the \ac{CB} for the final state.

The emission spectrum of the interband magnetoexciton is shown in Fig.~\ref{fig:Emission}, alongside spectra of the neutral exciton $X^0$ and the trion $X^-$ (magnetoexciton interacting with a single conduction electron) for reference (details on $X^0$ and $X^-$ are provided in \ac{SM}). Results are shown at $T=1$~mK [panel (a)] and $T=5$~K [panel (b)], with valley character indicated by color: red for $K'$ and blue for $K$.

The spectra of the magnetoexciton interacting with the $\nu=1$ state remain fully valley-polarized at all temperatures considered, unlike $X^0$, because the low-energy initial states are dominated by $K'$-valley configurations [Fig.~\ref{fig:Xcomparison}(a)]. A similar polarization occurs for $X^-$, where exchange interactions lower the energy of the $K'$-valley exciton when additional electrons occupy the same valley, producing a redshift relative to $X^0$; the redshift is smaller for $X^-$ due to only a single extra electron.  

The emission is intrinsically broadened even at low $T$, since the initial state is drawn from nearly degenerate low-energy configurations and the final state is an intraband exciton with an exceptionally flat spectrum [Fig.~\ref{fig:Xcomparison}(b)], where the lowest 20 states span $\sim2$~meV. Many-body interactions and this structure produce a broad emission profile, which is further widened at higher $T$ due to thermal occupation of excited initial states and additional recombination channels.

%%%%%%%%%%%%%%%%%%%%%%%%%%%%%%%%%%%%%%%%%%%%%%%%%%%%%%%%%%%%

%%%%%%%%%%%%%%%%%%%%%%%%%%%%%%%%%%%%%%
%\vspace{6pt}\textit{Conclusion}---
\vspace{6pt}\textit{Discussion}---
In this Letter we have developed a microscopic theory of the emission spectra of magnetoexcitons interacting with a spin- and valley-polarized gas of massive Dirac fermions in monolayer MoS$_2$ at $\nu=1$, where the lowest conduction-band Landau level is completely filled. By incorporating the magnetic field into the massive Dirac fermion model, establishing the corresponding optical selection rules, and performing exact diagonalization in the basis of electron–hole pair excitations, we obtained the correlated initial and final states governing the emission process. The resulting spectrum exhibits a pronounced redshift, strong broadening, and complete polarization, arising from exchange interactions with the polarized electron gas and the manifold of nearly degenerate excitonic configurations. The qualitative agreement with recent experimental data~\cite{Finley2021MoS2Magnetic, Finley2022MoS2Magnetic, Oreszczuk_2023} demonstrates that magneto-spectroscopy provides a sensitive probe of interacting massive Dirac fermions in atomically thin semiconductors.

%%%%%%%%%%%%%%%%%%%%%%%%%%%%%%%%%%%%%%%%%%%%%%%%%%%%%%%%%%%%
%%%%%%%%%%%%%%%%%%%%%%%%%%%%%%%%%%%%%%
\vspace{6pt}\textit{Acknowledgments}---
This work was supported by Mitacs through the Mitacs Globalink Research Award for research in Canada and by the Digital Research Alliance Canada with computing resources. P.H. and M.K. acknowledge financial support from the Quantum Sensors Challenge Program QSP078 and Applied Quantum Computing AQC004 Challenge Programs at the National Research Council of Canada, NSERC Discovery Grant No. RGPIN- 2019-05714, and University of Ottawa Research Chair in Quantum Theory of Materials, Nanostructures, and Devices. The authors thank J. J. Finley,  A. V. Stier and P. Kossacki for fruitful discussions.

%%%%%%%%%%%%%%%%%%%% Bibliography %%%%%%%%%%%%%%%%%%%%%%%%%%
%\bibliographystyle{apsrev4-2}
\bibliography{Bibliography}

\end{document}

\title{Supplementary Material for\\
``Magnetoexcitons and Massive Dirac Fermions in  Monolayers of Transition Metal Dichalcogenides in a High Magnetic Field''}

\maketitle

%\tableofcontents

%%%%%%%%%%%%%%%%%%%%%%%%%%%%%%%%%%%%%%
\section{Details on the \textit{ab initio}-based massive Dirac fermion model for a TMDC monolayer}
\label{section:DFT}
%%%%%%%%%%%%%%%%%%%%%%%%%%%%%%%%%%%%%%

%%%%%%%%%%%%%%%%% Fig %%%%%%%%%%%%%%%%
\begin{figure}[tb]
\includegraphics[width=0.6\linewidth]{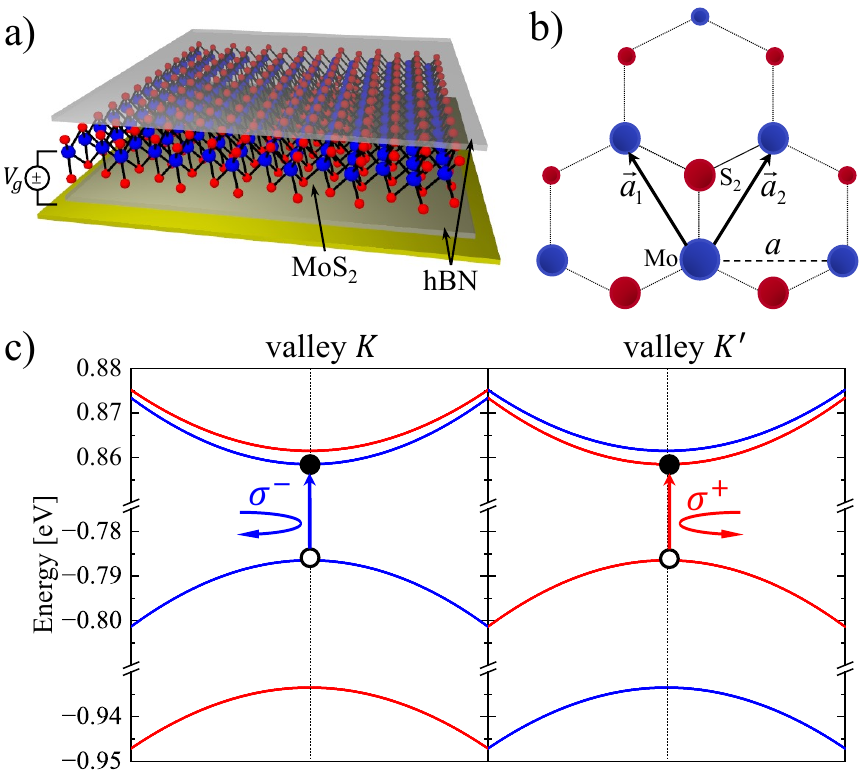}
    \caption{
    Geometry and electronic properties of a  MoS$_2$ monolayer. 
    (a) Schematic representation of the \ac{3D} device layout, showing the MoS$_2$ monolayer (blue and red spheres) encapsulated by \ac{hBN}, with a metallic gate (yellow surface) that enables tuning of the carrier density via electrostatic doping.
    (b) Top view of the single-layer \ac{TMDC} honeycomb lattice. Metal atoms are represented in dark blue, while chalcogen atoms are shown in red. The in-plane lattice constant is denoted as $a$. The lattice vectors are chosen as $\vec{a}_{1}=\left(-a/2,a\sqrt{3}/2\right)$ and $\vec{a}_2=\left(a/2,a\sqrt{3}/2\right)$, respectively.
    (c) The electronic band structure of monolayer MoS$_2$ within the \ac{mDF} model, including \ac{SOC}, shown for both the $K$ and $K'$ valleys. The spin projections of the bands are indicated by color, where red/blue denotes spin up ($\uparrow$)/down ($\downarrow$), respectively. Allowed interband transitions under circularly polarized light are also shown, with $\sigma^\pm$ denoting the light polarization.}
\label{fig:TMDCgeometry}
\end{figure}
%%%%%%%%%%%%%%%%%%%%%%%%%%%%%%%%%%%%%%

%The electronic  properties of monolayer MoS$_2$ are obtained using first-principles computations.

The atomic structure of a representative monolayer \ac{TMDC}, MoS$_2$, is shown in Fig.~\ref{fig:TMDCgeometry}, where panel (a) reflects a typical experimental configuration similar to Ref.~\cite{Finley2021MoS2Magnetic}. We consider a single \ac{TMDC} layer in the $2H$ phase, encapsulated by \ac{hBN} (gray layers), as explicitly shown in the figure.  A metallic gate, represented by the yellow surface beneath the atomic layers, allows to tune the carrier density via electrostatic doping.

Structurally, the $2H$ phase exhibits a hexagonal crystal lattice, with a lattice constant $a = 3.160$~\AA, corresponding to the distance between neighboring molybdenum (Mo) atoms, as shown in Fig.~\ref{fig:TMDCgeometry}(b). Each Mo atom is surrounded in-plane by three \ac{NN} sulfur dimers (S$_2$), arranged at a distance of $a\sqrt{3}/3 = 1.82$~\AA, forming the characteristic honeycomb pattern observed in the top view of the monolayer.
The primitive lattice vectors in real space are given by $\vec{a}_{1} = \left(-a/2,\, a\sqrt{3}/2\right)$ and $\vec{a}_2 = \left(a/2,\, a\sqrt{3}/2\right)$. The vertical separation between the top and bottom chalcogen layers is $3.17$~\AA, while the metal–chalcogen bond length is $2.42$~\AA.

% First principles electronic structure  of a MoS$_2$ monolayer
The single-particle electronic structure of ungated monolayer MoS$_2$ is obtained using \ac{DFT} calculations, following the workflow outlined in Refs.~\cite{bieniek2022nanomaterials,Sadecka2025heteroQD}. At zero applied vertical electric field, MoS$_2$ exhibits \ac{CB} minima and \ac{VB} maxima at the $K$ and $K'$ points of the Brillouin zone, leading to a direct $K$--$K$ band gap, in contrast to the indirect nature of the multi-layer (bulk) material~\cite{He2014stacking}. The calculated band gap is $\Delta = 1.27$~eV. The \ac{SOC} introduces spin splittings of $\Delta_{\text{SOC}}^{\text{CB}} = 2.7$~meV and $\Delta_{\text{SOC}}^{\text{VB}} = 147$~meV in the \ac{CB} and \ac{VB}, respectively. In each valley, the spin projections of the highest-energy \ac{VB} subband and the lowest-energy \ac{CB} subband are the same, which enables radiative recombination of the lowest-energy electron–hole pair and thus makes the optical transition spin-allowed (bright), even in the presence of \ac{SOC}. The spin character of the band edges, combined with valley-selective optical selection rules, results in spin–valley locking near the $K$ points, very strong in the \ac{VB} and weaker in the \ac{CB}. Importantly, both band extrema are dominated by Mo $d$ orbitals, the \ac{CB} mainly by $4d_{m=0}$ and the \ac{VB} by $4d_{m=\pm2}$, with the  sign of the latter dependent on the valley.

% Massive Dirac Fermions model and the inclusion of a perpendicular magnetic field
The \ac{mDF} model is parametrized by the band gap, Fermi velocity, and spin–orbit splittings extracted from the \ac{DFT} calculations. This model captures the essential low-energy features of monolayers of \acp{TMDC}, including MoS$_2$---the direct band gap, spin–valley coupling~\cite{bradlyn2016beyond}, and the orbital composition of the band edges~\cite{Szulakowska2019magnetoX}---while providing a computationally tractable platform for modeling the optical properties of \acp{TMDC} in external magnetic field and varying charge densities~\cite{xiao2012coupled,Kormanyos2013,VanDerDonck2018}. 

We start from the effective low-energy \ac{mDF} Hamiltonian~\cite{Xiao2012,Kormanyos2013,VanDerDonck2018}:
%%%%%%%%%%%%%%%%%%%%%%%%%%%%%
\begin{equation}
    \hat{H}_{s,\tau}(\vec{k}) = 
    at\sqrt{3}/2\left(\tau k_x\sigma_x + k_y\sigma_y\right) 
    + \frac{\Delta}{2}\sigma_z 
    + \frac{\lambda s\tau}{2}\left(I_2 - \sigma_z\right),
    \label{eq:mDF_Hgeneral}
\end{equation}
%%%%%%%%%%%%%%%%%%%%%%%%%%%%%
where $\sigma_i$ ($i = x, y, z$) are the Pauli matrices acting in the $\{\text{CB}, \text{VB}\}$ basis, $I_2$ is the $2 \times 2$ identity matrix, $a$ is the lattice constant, $t$ is the hopping integral, $\tau = \pm 1$ labels the $K$ and $K'$ valley, $s = \pm 1$ denotes the spin, and $\lambda$ is the \ac{SOC} parameter, to be specified later on.

The \ac{mDF} model was developed originally for systems such as the monolayer \ac{hBN}, in which the hexagonal lattice is composed of two interpenetrating triangular sublattices, one built out of atoms B, and the other of atoms N, with one  orbital per atom. In such material, the basis of the Hamiltonian consists of the two Bloch wavefunctions, each accounting for one sublattice. The band states in the vicinity of the points $K/K'$ can therefore be mapped onto the surface of the pseudospin Bloch sphere, with pseudospin up and down states idenitified respectively with sublattice B and N. For MoS$_2$, on the other hand, the model accounting for one orbital per atom is not valid, since at points $K/K'$ both Hamiltonian eigenstates are built predominantly out of even $d$ orbitals of the Mo atom, as described in the previous Section ($d_{m=0}$ for the \ac{CB} and $d_{m=\pm 2}$ for the \ac{VB}). The band states in the vicinity of these points are therefore superpositions of all even Mo orbitals, with contributions from S atoms present to a much smaller degree. The microscopic nature of the band states is therefore much more complex than that in \ac{hBN} and cannot be fully expressed in the sublattice index (pseudospin) notation.

Let us first consider the first two terms of the Hamiltonian ~\eqref{eq:mDF_Hgeneral}. By defining the Fermi velocity $v_F = -t a \sqrt{3}/2$ and expanding around the valley centers $\vec{k} = \vec{K} + \vec{q}$, we can rewrite the Hamiltonian as~\cite{geim2013van}:
%%%%%%%%%%%%%%%%%%%%%%%%%%%%%
\begin{equation}
    \hat{H}_{\tau}(\vec{k}) = 
    \begin{bmatrix}
        \frac{\Delta}{2} & v_F \left(\tau q_x - i q_y \right) \\
        v_F \left(\tau q_x + i q_y \right) & -\frac{\Delta}{2}
    \end{bmatrix},
    \label{eq:MagnetoX_mDF_H}
\end{equation}
%%%%%%%%%%%%%%%%%%%%%%%%%%%%%
where the hopping integral $t = -1.4677$~eV has been fitted to match the curvature of the \ac{DFT} bands, and $\Delta = 1.27$~eV is the direct band gap from \ac{DFT}. 
 
Next, let us include the \ac{SOC} term. In the spin-resolved basis \mbox{$\{\text{CB}\downarrow,~\text{VB}\downarrow,~\text{CB}\uparrow,~\text{VB}\uparrow\}$}, the Hamiltonian for the valley $\tau$ takes the form: 
%%%%%%%%%%%%%%%%%%%%%%%%%%%%%
\begin{equation}
\begin{split}
&\hat{H}_{\tau}(\vec{k}) = \\
&\left[
\begin{array}{@{\hskip 0pt}c@{\hskip 2pt}c@{\hskip 2pt}c@{\hskip 2pt}c@{\hskip 0pt}}  % adjust spacing here
    \frac{\Delta}{2} \!+\! \tau \frac{\Delta_{\text{SOC}}^{\text{CB}}}{2} & v_F ( \tau q_x \!-\! i q_y ) & 0 & 0 \\
    v_F ( \tau q_x \!+\! i q_y ) & -\frac{\Delta}{2} \!-\! \tau \frac{\Delta_{\text{SOC}}^{\text{VB}}}{2} & 0 & 0 \\
    0 & 0 & \frac{\Delta}{2} \!-\! \tau \frac{\Delta_{\text{SOC}}^{\text{CB}}}{2} & v_F ( \tau q_x \!-\! i q_y ) \\
    0 & 0 & v_F ( \tau q_x \!+\! i q_y ) & -\frac{\Delta}{2} \!+\! \tau\frac{\Delta_{\text{SOC}}^{\text{VB}}}{2}
\end{array}
\right]
\end{split}
\label{eq:MagnetoX_mDF_H_SOC}
\end{equation}
%%%%%%%%%%%%%%%%%%%%%%%%%%%%%
where $\Delta^{\text{CB}}_{\text{SOC}} = -3$~meV and $\Delta^{\text{VB}}_{\text{SOC}} = 147$~meV, as extracted from the \ac{DFT} results. This form explicitly reflects the  spin--valley locking around the $K$ and $K'$ valleys, which is particularly strong for the \ac{VB}.

The spin-resolved energy structure resulting from the diagonalization of the Hamiltonian given in Eq.~\eqref{eq:MagnetoX_mDF_H_SOC} is illustrated in Fig.~\ref{fig:TMDCgeometry}(c). The two-band model for each spin gives rise to spin-split \ac{CB} and \ac{VB} in both $K$ and $K'$ valleys, with opposite ordering due to time-reversal symmetry. The spin projection is indicated by color, and the energy splittings are consistent with the \ac{DFT}-derived values of $\Delta^{\text{CB}}_{\text{SOC}}$ and $\Delta^{\text{VB}}_{\text{SOC}}$. While a detailed discussion of light–matter coupling is deferred to the next Section, Fig.~\ref{fig:TMDCgeometry}(c) also schematically indicates the valley-selective optical transitions allowed by circularly polarized light ($\sigma^\pm$), reflecting the spin–valley–polarization coupling intrinsic to monolayer MoS$_2$.

%%%%%%%%%%%%%%%%%%%%% Magnetic Field Effect  %%%%%%%%%%%%%%%%%%%%%%%%%%%%%%%%%%%%%%%%%%%
The external magnetic field is incorporated by replacing the momentum operator $\vec{p} = -i\hbar\nabla$ in Eq.~\eqref{eq:MagnetoX_mDF_H} with the generalized momentum $\vec{q} \to \vec{p} + |e|\vec{A}$. A magnetic field oriented along the $z$-axis, $\vec{B} = [0, 0, B]$, is considered, along with the symmetric gauge for the vector potential, $\vec{A} = [-By/2, Bx/2, 0]$. Next, the corresponding ladder operators are introduced:
%%%%%%%%%%%% Eq %%%%%%%%%%%%%
\begin{equation}
\begin{split}
    \hat{a} &= \frac{l_B}{\sqrt{2}} \left(\partial_x-i\partial_y+\frac{1}{2l_B^2}(-iy+x)\right) \\
    \hat{a}^\dagger &= \frac{l_B}{\sqrt{2}} \left(-\partial_x-i\partial_y+\frac{1}{2l_B^2}(iy+x)\right)
\end{split},
\label{eq:MagnetoX_mDF_B_ladder}
\end{equation}
%%%%%%%%%%%%%%%%%%%%%%%%%%%%%
where $l_B = \sqrt{\hbar/eB}$ denotes the magnetic length. These operators satisfy the commutation relation $[\hat{a}, \hat{a}^\dagger] = 1$. By introducing the parameter $c = v_F \sqrt{2}/l_B$, the Hamiltonian can be expressed, for each valley, in the basis of spin-resolved low-energy bands, \mbox{$\{\text{CB}\downarrow,\text{VB}\downarrow,\text{CB}\uparrow,\text{VB}\uparrow\}$}, as:
%%%%%%%%%%%%%%%%%%%%%%%%%%%%%
\begin{widetext}
\allowdisplaybreaks{
\begin{align}
    \hat{H}_B(K) &= 
    \begin{bmatrix}
        \frac{\Delta}{2}+\frac{\Delta^{\text{CB}}_{\text{SOC}}}{2}-\frac{g_e\mu_B B}{2} & -ic\hat{a} & 0 & 0 \\
        ic \hat{a}^\dagger & -\frac{\Delta}{2}-\frac{\Delta^{\text{VB}}_{\text{SOC}}}{2}-\frac{g_h\mu_B B}{2} & 0 & 0 \\
        0 & 0 & \frac{\Delta}{2}-\frac{\Delta^{\text{CB}}_{\text{SOC}}}{2}+\frac{g_e\mu_B B}{2} & -ic \hat{a} \\
        0 & 0 & ic \hat{a}^\dagger & -\frac{\Delta}{2}+\frac{\Delta^{\text{VB}}_{\text{SOC}}}{2}+\frac{g_h\mu_B B}{2}
    \end{bmatrix} \nonumber \\ 
    \hat{H}_B(K') &= 
    \begin{bmatrix}
        \frac{\Delta}{2}+\frac{\Delta^{\text{CB}}_{\text{SOC}}}{2}-\frac{g_e\mu_B B}{2} & -ic \hat{a}^\dagger & 0 & 0 \\
        ic \hat{a} & -\frac{\Delta}{2}-\frac{\Delta^{\text{VB}}_{\text{SOC}}}{2}-\frac{g_h\mu_B B}{2} & 0 & 0 \\ \vspace{3mm}
        0 & 0 & \frac{\Delta}{2}-\frac{\Delta^{\text{CB}}_{\text{SOC}}}{2}+\frac{g_e\mu_B B}{2} & -ic \hat{a}^\dagger \\
        0 & 0 & ic \hat{a} & -\frac{\Delta}{2}+\frac{\Delta^{\text{VB}}_{\text{SOC}}}{2}+\frac{g_h\mu_B B}{2}
    \end{bmatrix},
\label{eq:mDFmagneticH}
\end{align}}
\end{widetext}
%%%%%%%%%%%%%%%%%%%%%%%%%%%%%
where $\mu_B=e\hbar/2m_e$ is the Bohr magneton, and $g_e$ and $g_h$ are the electron and hole $g$ factors. In the following we take $g_e=-2$ and $g_h=-2.2$~\cite{Stier2021PRR}.

The magnetic field enters the Hamiltonian through two effects. First, the Zeeman effect modifies the diagonal terms via subband-dependent $g$-factors, leading to energy shifts that scale linearly with $B$. Second, the off-diagonal terms include the magnetic field through the orbital (diamagnetic) effect, which leads to the formation of discrete \ac{LL}s. 

The eigenenergies (\acp{LL}) and wavefunctions for each valley can be obtained analytically by diagonalizing $\hat{H}_B(K)$ and $\hat{H}_B(K')$, respectively. 
The wavefunctions for each valley can be expressed as:
%%%%%%%%%%%% Eq %%%%%%%%%%%%%
\begin{equation}
\begin{split}
    &\ket{\psi^{\text{VB/CB}}_{n,m}(K)} = 
    \begin{pmatrix}
        \alpha^{\text{VB/CB}}_{n} \ket{n-1,m} \\
        \beta^{\text{VB/CB}}_{n} \ket{n,m}
    \end{pmatrix}
    \\
    &\ket{\psi^{\text{VB/CB}}_{n,m}(K')} = 
    \begin{pmatrix}
        \alpha^{\text{VB/CB}}_{n} \ket{n,m} \\
        \beta^{\text{VB/CB}}_{n} \ket{n-1,m}
    \end{pmatrix}.
\end{split}
\label{eq:mDFmagnetic_wf}
\end{equation}
%%%%%%%%%%%%%%%%%%%%%%%%%%%%%
These wavefunctions are subband spinors expressed in terms of the eigenstates $|n,m\rangle$ of the \ac{2D} \ac{HO}. The ladder operators $\hat{a}$, $\hat{a}^{\dagger}$ are associated with the quantum number $n$. In our \ac{2D} system there exists another pair of chiral \ac{HO} operators $\hat{b}$, $\hat{b}^{\dagger}$, commuting with $\hat{a}$, $\hat{a}^{\dagger}$, and associated with the quantum number $m$. Since the Hamiltonians $\hat{H}_B(K)$ and $\hat{H}_B(K')$ do not contain the operators $\hat{b}$, $\hat{b}^{\dagger}$, the mDF eigenenergies do not depend on $m$, resulting in macroscopic  degeneracies of \acp{LL}.

\begin{figure}[tb]
\includegraphics[width=0.6\linewidth]{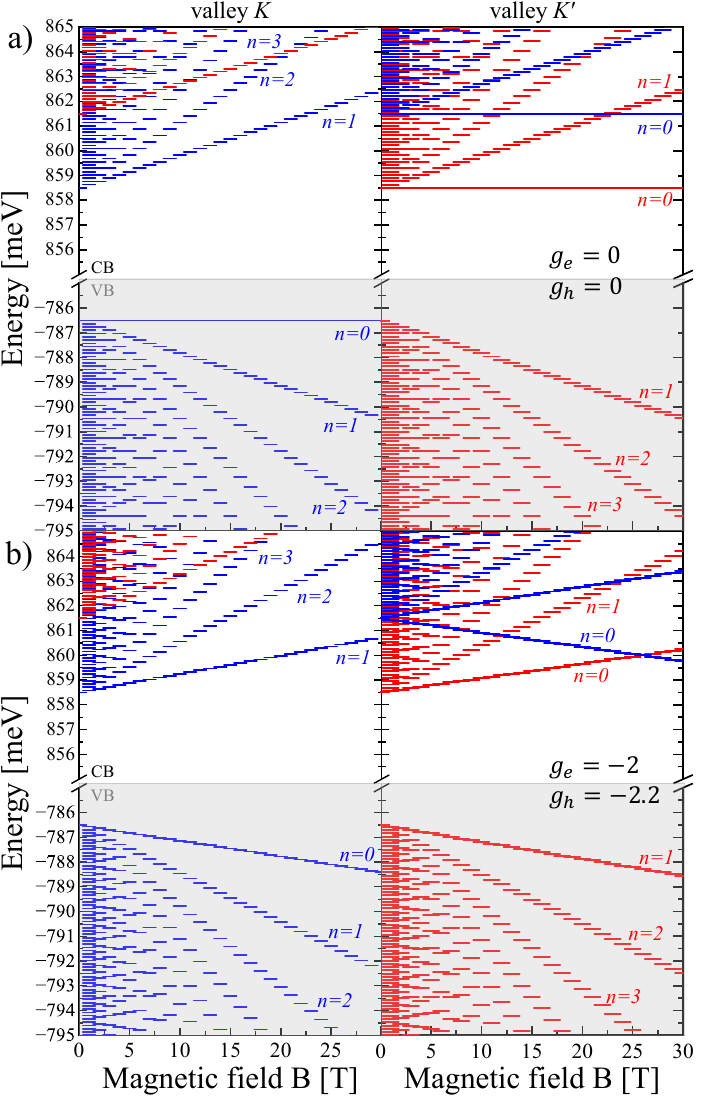}
    \caption{Spin- and valley-resolved \acp{LL} spectrum as a function of the vertical magnetic field for a MoS$_2$ monolayer (a) with the $g$-factors set to zero and (b) with the non-zero $g$ factors appropriate for MoS$_2$.}
\label{fig:LLvsB}
\end{figure}
%%%%%%%%%%%%%%%%%%%%%%%%%%%%%%%%%%%%%%

As can be deduced from the structure of the band spinors given in Eq.~\eqref{eq:mDFmagnetic_wf}, the case of $n=0$ is special in that the eigenstates are reduced to one component only (the second in valley $K$ and the first in valley $K'$). These cases constitute the zeroth \acp{LL}, whose energies originate solely from the diagonal terms in the Hamiltonian and remain unaffected by the off-diagonal (diamagnetic) coupling $c \propto \sqrt{B}$. The \ac{LL} fan diagram, shown in Fig.~\ref{fig:LLvsB}(a), has been constructed for the $K$ (left panel) and $K'$ (right panel) valleys under the assumption $g_e = g_h = 0$. Levels with $n > 0$ exhibit an approximately linear dependence on $B$, while the $n = 0$ levels in both valleys remain constant with respect to the magnetic field. This behavior is a characteristic of Dirac-like systems, including monolayer and bilayer graphene~\cite{neto2009electronic, McCann_prl2006}, and reflects the chiral symmetry inherent in the Hamiltonian. In Fig.~\ref{fig:LLvsB}(b), the fan diagrams are shown with Zeeman terms included. The resulting spin-dependent energy shifts lead to tilted \acp{LL} and enhanced valley splitting, especially at higher magnetic fields. The strong spin--valley locking in the \ac{VB}, driven by the large spin--orbit interaction, gives rise to valley-specific bright and dark excitonic states, depending on the spin orientation. This feature directly influences the polarization and intensity of optical transitions. Since the primary objective is to explore strongly correlated electrons through magneto-optical spectroscopy, the spin-polarized character of the valence states becomes central for interpreting the emerging many-body phenomena in the quantum Hall regime. In contrast, the \ac{CB}, shaped by a relatively weaker \ac{SOC}, hosts a dense manifold of intersecting \acp{LL} originating from various spin-split subbands. This results in a complex pattern of level crossings and reorderings as the magnetic field increases. The evolution of inter-\ac{LL} energy gaps with magnetic field produces a nontrivial sequence of \ac{LL} filling as a function of increasing electron density. In the absence of electron--electron interactions, the lowest spin-split conduction level in one valley is filled first, followed by the corresponding level in the opposite valley.

%%%%%%%%%%%%%%%%%%%%%%%%%%%%%%%%%%%%%%
\section{Interband optical selection rules and transition rates in the mDF model in the presence of a magnetic field}
\label{section:SelectionRules}
%%%%%%%%%%%%%%%%%%%%%%%%%%%%%%%%%%%%%%

In the following Section we discuss the selection rules for optical transitions connecting the single-particle orbitals. The MoS$_2$ monolayer is illuminated by a monochromatic electromagnetic wave incident perpendicularly, with circular polarization $\sigma^{\pm}$ and frequency $\omega$. The electric field components are given by 
$E_x = \pm E_0 \sin{\omega t}$ and
$E_y = E_0 \cos{\omega t}$,
where $E_0$ denotes the field amplitude. Within the momentum-space formalism expressed in Eq.~\eqref{eq:MagnetoX_mDF_H_SOC}, the electromagnetic field is introduced through the time-dependent vector potential,
$A_x = \pm \frac{E_0}{\omega} \cos{(\omega t)}$ and 
$A_y = -\frac{E_0}{\omega} \sin{(\omega t)}$,
by modifying the generalized momentum operator, following the same approach used for the static magnetic field~\cite{trushin2016optical, Szulakowska2019magnetoX}. The light field is treated as a weak perturbation, and resulting excitations are assumed to conserve both spin and valley indices. In the basis \mbox{$\{\text{CB},s,~\text{VB},s\}$} corresponding to the Hamiltonians in Eq.~\eqref{eq:mDFmagneticH}, the interaction between the electron system and the light field in valley $\tau$ is described by the following Hamiltonian:
%%%%%%%%%%%%%%%%%%%%%%%%%%%%%%%%%%%%
\begin{equation}
\hat{H}'(\tau,\sigma^\pm) =\pm \tau g 
\begin{bmatrix}
    	0 & e^{\pm \tau i \omega t} \\
    	e^{\mp \tau i \omega t} & 0
    \end{bmatrix},
\label{eq:emHamil}
\end{equation}
where $g = v_F e E_0 / \omega$.
%%%%%%%%%%%%%%%%%%%%%%%%%%%%%%%%%%%%

In the next step, the Hamiltonians in Eq.~\eqref{eq:mDFmagneticH} are diagonalized in the absence of illumination, yielding the \ac{LL} energies $E_n^\text{CB}$ and $E_n^\text{VB}$ for the conduction and valence bands, respectively, along with the corresponding spinor wavefunctions given in Eq.~\eqref{eq:mDFmagnetic_wf}. It is assumed that at time $t=0$ the electron occupies a \ac{LL} in the \ac{VB}, and the illumination is then switched on instantaneously. Consequently, the time evolution of the system in valley $\tau$ is described by the wavefunction:
%%%%%%%%%%%%%%%%%%%%%%%%%%%%%%%%%%%%
\begin{equation}
    \ket{\psi_{\tau,n,m,n',m'}^{\text{B}}(t)} =
    C^{\text{VB}}_{\tau,n,m}(t)e^{-\frac{i}{\hbar}E_{n,m}^{\text{VB}}t}\ket{\psi^{\text{VB}}_{\tau,n,m}} + 
    C^{\text{CB}}_{\tau,n',m'}(t)e^{-\frac{i}{\hbar}E_{n',m'}^{\text{CB}}t}\ket{\psi^{\text{CB}}_{\tau,n',m'}},
\end{equation}
%%%%%%%%%%%%%%%%%%%%%%%%%%%%%%%%%%%%
with initial conditions
$C^{\text{VB}}_{\tau,n,m}(t=0)=1$ and
$C^{\text{CB}}_{\tau,n',m'}(t=0)=0$.
We note that the above superposition can contain single-particle wavefunctions belonging to \acp{LL} labeled by {\em different} indices $n$. The coefficients $C^{\text{VB/CB}}_{\tau,n,m}(t=0)$ are obtained by solving the time-dependent Schr\"odinger equation:
%%%%%%%%%%%%%%%%%%%%%%%%%%%%%%%%%%%%
\begin{equation}
    i\hbar \frac{\partial}{\partial t} \ket{\psi_{\tau{,n,m,n',m'}}^{\text{B}}(t)} 
    = \left( \hat{H}_{\text{B}}(\tau) + \hat{H}'(\tau,\sigma^\pm) \right) \ket{\psi_{\tau{,n,m,n',m'}}^{\text{B}}(t)},
\label{eq:MagnetoX_TDSE_LL}
\end{equation}
%%%%%%%%%%%%%%%%%%%%%%%%%%%%%%%%%%%%
accounting for the complete system Hamiltonian,
$\hat{H}_{\text{\text{B}}}(\tau) + \hat{H}'(\tau,\sigma^\pm)$, composed of the Hamiltonians \eqref{eq:mDFmagneticH} and \eqref{eq:emHamil}.

To proceed further, the Hamiltonian $\hat{H}'(\tau,\sigma^{\pm})$ is rotated into the basis of the \ac{LL} spinors. We consider only the interband optical selection rules (transitions between the valence and conduction bands). The expression for the valley $K$ is obtained as:
%%%%%%%%%%%%%%%%%%%%%%%%%%%%%%%%%%%%
\begin{eqnarray}
    &\hat{H}^\text{CV}_{K,n,m,n',m'}=\braket{ \psi^{\text{CB}}_{K,n',m'} | \hat{H}'(\sigma^\pm) | \psi^{\text{VB}}_{K,n,m} } = \nonumber \\
    &\pm g (\alpha_{{n'}}^{\text{CB}})^* \beta_{{n}}^{\text{VB}} e^{\pm i\omega t} {\langle n'-1,m' | n,m \rangle} 
    \pm g \alpha_{{n}}^{\text{VB}} (\beta_{{n'}}^{\text{CB}})^* e^{\mp i\omega t} {\langle n',m' | n-1,m \rangle}, \nonumber \\
\end{eqnarray}
%%%%%%%%%%%%%%%%%%%%%%%%%%%%%%%%%%%%
Due to the orthogonality of the \ac{HO} orbitals, the diagonal elements of this Hamiltonian vanish, but the off-diagonal elements can be finite under conditions discussed below. Analogous formula can be derived for the valley $K'$.

For short times $t$, the approximation $C^{\text{VB}}_{\tau,n,m}(t) \approx 1$ can be applied (corresponding to a series expansion to lowest order in $E_0$), allowing the focus to be placed on the coefficients $C^{\text{CB}}_{\tau,n',m'}(t)$. With $\Delta E = E^{\text{CB}}_{n'} - E^{\text{VB}}_{n}$, the equation of motion governing these coefficients are given by:
%%%%%%%%%%%%%%%%%%%%%%%%%%%%%%%%%%%%%%%%%%%%%%%%%%%%
\begin{equation}
    \frac{\partial}{\partial t}C_{\tau,n',m'}^{\text{CB}}(t) =
    C_{\tau,n,m}^{\text{VB}}(t) e^{\frac{i}{\hbar}\Delta E t} \hat{H}^\text{CV}_{\tau,n,m,n',m'} 
\end{equation}
%%%%%%%%%%%%%%%%%%%%%%%%%%%%%%%%%%%%%%%%%%%%%%%%%%%%
which for each valley and polarization can be expressed as:
%%%%%%%%%%%%%%%%%%%%%%%%%%%%%%%%%%%%%%%%%%%%%%%%%%%%
\begin{equation}
\begin{split}
    &\frac{\partial}{\partial t}C_{K,n',m'}^{\text{CB}}(t) = 
    \mp\frac{ig}{\hbar}  
    \left[ 
    \begin{split}
        &e^{i\left(\mp\omega+\frac{\Delta E}{\hbar}\right)t} \alpha^{\text{VB}}_{{n'}}(\beta^{\text{CB}}_{{n}})^* {\braket{n',m'|n-1,m}} + \\
        &e^{i\left(\pm\omega+\frac{\Delta E}{\hbar}\right)t} \beta^{\text{VB}}_{{n'}}(\alpha^{\text{CB}}_{{n}})^* {\braket{n'-1,m'|n,m}}
    \end{split}
    \right], \\
    &\frac{\partial}{\partial t}C_{K',n',m'}^{\text{CB}}(t) = \pm\frac{ig}{\hbar}  
    \left[ 
    \begin{split}
        &e^{i\left(\pm\omega+\frac{\Delta E}{\hbar}\right)t} \alpha^{\text{VB}}_{{n'}}(\beta^{\text{CB}}_{{n}})^* {\braket{n'-1,m'|n,m}} + \\
        &e^{i\left(\mp\omega+\frac{\Delta E}{\hbar}\right)t} \beta^{\text{VB}}_{{n'}}(\alpha^{\text{CB}}_{{n}})^* {\braket{n',m'|n-1,m}}
    \end{split}
    \right].
\end{split}
\label{eq:MagnetoX_TDSE_cCB_LL}
\end{equation}
%%%%%%%%%%%%%%%%%%%%%%%%%%%%%%%%%%%%%%%%%%%%%%%%%%%%
The terms enclosed in square brackets represent the various allowed transitions between the \acp{LL}, encompassing the couplings between valence and \ac{CB} states as well as the polarization dependence of the light field. It is noted that the \ac{CB} spinor is dominated by the conduction band component, i.e., $\alpha^{\text{CB}}_{n'} \gg \beta^{\text{CB}}_{n'}$, whereas the \ac{VB} spinor exhibits the opposite character. This asymmetry permits the neglect of the first term in each square bracket without compromising the generality of the approach. Additionally, contributions in which the frequency $\omega$ appears with a negative sign exhibit rapid temporal oscillations and thus average out to zero over time. As a result, the equations of motion reduce to a single expression for each valley and light polarization:
%%%%%%%%%%%%%%%%%%%%%%%%%%%%%%%%%%%%%%%%%%%%%%%%%%%%
\begin{eqnarray}
 (K,\sigma^-) &:& 
    \frac{\partial}{\partial t}C_{n',m'}^{\text{CB}}(t) = \frac{ig}{\hbar}  
    e^{i\left(-\omega+\frac{\Delta E}{\hbar}\right)t} \beta^{\text{VB}}_{{n}}(\alpha^{\text{CB}}_{{n'}})^* {\braket{n'-1,m'|n,m}}, \nonumber \\
(K',\sigma^+) &:& 
    \frac{\partial}{\partial t}C_{n',m'}^{\text{CB}}(t) = \frac{ig}{\hbar}  
    e^{i\left(-\omega+\frac{\Delta E}{\hbar}\right)t} \beta^{\text{VB}}_{{n}}(\alpha^{\text{CB}}_{{n'}})^* {\braket{n',m'|n-1,m}}. \nonumber
\label{eq:MagnetoX_mDF_LightCoupling_LL_full}
\end{eqnarray}
%%%%%%%%%%%%%%%%%%%%%%%%%%%%%%%%%%%%%%%%%%%%%%%%%%%%
Each valley has its own selection rules for the allowed transitions, determined by the polarization of the light. Moreover, the projections of \ac{HO} orbitals define the optical selection rules: in the valley $K$, the transition from the \ac{VB} \ac{LL} (unprimed indices) into the \ac{CB} \ac{LL} (primed indices) requires $m'=m$ and $n'=n+1$, i.e., the \ac{LL} index has to be augmented by one unit, whilen in the valley $K'$ we have $m'=m$ and $n'=n-1$, i.e., the \ac{LL} index has to be decreased by one unit.
To simplify the notation, the dipole matrix elements are defined in the following form
%%%%%%%%%%%%%%%%%%%%%%%%%%%%%%%%%%%%%%%%%%%%%%%%%%%%
\begin{equation}
    D_{n',n} = \frac{g}{\hbar} \beta^{\text{VB}}_{n}(\alpha^{\text{CB}}_{n'})^*.
\end{equation}
%%%%%%%%%%%%%%%%%%%%%%%%%%%%%%%%%%%%%%%%%%%%%%%%%%%%

The valley- and polarization-resolved transition rates are obtained by integrating the above equations of motion. In the long-time limit, this gives:
%%%%%%%%%%%%%%%%%%%%%%%%%%%%%%%%%%%%%%%%%%%%%%%%%%%%
\begin{align}
    &(\sigma^-, K): \gamma_{n,m,n',m'} = |C^{\text{CB}}_{K,n',m'}|^2_{t\rightarrow\infty} = 2\pi \left| D_{n',n} \right|^2 \delta \Bigl( \omega - \frac{\Delta E}{\hbar} \Bigl) |\braket{n'-1,m'|n,m}|^2, \\
    &(\sigma^+, K'): \gamma_{n,m,n',m'} = |C^{\text{CB}}_{K',n',m'}|^2_{t\rightarrow\infty} = 2\pi \left| D_{n',n}\right|^2 \delta \Bigl( \omega - \frac{\Delta E}{\hbar} \Bigl)  |\braket{n',m'|n-1,m}|^2.
\end{align}
%%%%%%%%%%%%%%%%%%%%%%%%%%%%%%%%%%%%%%%%%%%%%%%%%%%%
This result is directly related to Fermi's Golden Rule, where the transition rate is governed by the matrix elements of the perturbation and the energy conservation condition, represented by the delta function.

As is evident from this discussion, the transition rates between the \ac{VB} and \ac{CB} \acp{LL} depend on the specific \ac{LL} indices $n$, $n'$, as the values of the spinor coefficients $\alpha$ and $\beta$ depend on these indices. However, the optical selection rules emerge clearly for any pair of the \ac{LL} states
$(n,m)$ (in the \ac{VB}) and $(n',m')$ (in the \ac{CB}):
%%%%%%%%%%%%%%%%%%%%%%%%%%%%%%%%%%%%%%%%%%%%%%%%%%%%
\begin{equation}
\begin{split}
    (\sigma^-, K) &:  \hspace{6mm} n' - n = +1,  \hspace{3mm} m' = m, \\
    (\sigma^+, K') &: \hspace{6mm} n' - n = -1,  \hspace{3mm} m' = m.
\end{split}
\label{eq:MagnetoX_mDF_LightCoupling_LL}
\end{equation}
%%%%%%%%%%%%%%%%%%%%%%%%%%%%%%%%%%%%%%%%%%%%%%%%%%%%
In these selection rules, the  quantum number $m$ (the index of an orbital within a \ac{LL}) remains unchanged, while the \ac{LL} index $n$ undergoes a shift by one unit for each valley. The $\sigma^-$ polarization induces a transition where the $K$-valley \ac{LL} index increases by one ($n' = n + 1$), while the $K'$ valley transitions are forbidden.
The $\sigma^+$ polarization causes a transition where the \ac{LL} index decreases by one ($n' = n - 1$), and such transition can only take place in the valley $K'$. The valley-resolved selection rules arise from the breaking of inversion and time-reversal symmetries in the presence of a magnetic field, demonstrating how optical activity is distributed between the $K$ and $K'$ valleys in monolayer MoS$_2$.
 
%The allowed transitions are schematically shown in Fig.~\ref{fig:TMDCwithLL} as vertical arrows connecting specific \acp{LL}. These transitions conserve the \ac{LL} number $ m $ (which is not directly visible in the Figure) and involve changes in the \ac{LL} index $ n $ by $\pm 1$. The valley-resolved selection rules arise from the breaking of inversion and time-reversal symmetries in the presence of a magnetic field, demonstrating how optical activity is distributed between the $K$ and $K'$ valleys in monolayer MoS$_2$.

%%%%%%%%%%%%%%%%%%%%%%%%%%%%
\section{Coulomb interaction matrix elements}
\label{section:AppendixCME}
%%%%%%%%%%%%%%%%%%%%%%%%%%%%

In this Section, the procedure for computing the matrix elements of the Coulomb interaction, essential for the numerical solution of the \ac{BSE} in the charged magnetoexciton problem, is outlined. The Coulomb potential, $V_C(\vec{r}_1,\vec{r}_2) = \frac{e^2}{4\pi\varepsilon_0\varepsilon} \frac{1}{|\vec{r}_1-\vec{r}_2|}$, governs the scattering between two electrons, where $e$ denotes the elementary charge and $\varepsilon_0$ and $\varepsilon$ are the vacuum and relative dielectric constants, respectively. The required matrix elements take the form $\braket{i,j|V_C|k,l}$, with the composite indices $i=\{n,m,\tau,s\}$ (similarly for $j$, $k$, and $l$) labeling the \ac{LL} spinors defined in Eq.~\eqref{eq:mDFmagnetic_wf}. In this notation, $n$ and $m$ correspond to the \ac{LL} index and the intra-level quantum number, respectively, while $\tau$ and $s$ denote the valley and spin indices. The outer (inner) indices specify the electron located at position $\vec{r}_1$ ($\vec{r}_2$).

To elucidate the details of the derivation, the \ac{LL} orbitals, Eq.~(\ref{eq:mDFmagnetic_wf}), are expressed in the real-space representation. For the valley $K$, the corresponding wavefunction takes the form:
%%%%%%%%%%%%%%%%%%%%%%%%%%%%%%%%%%%%%%%%%%%%%%%%%%%%
\begin{equation}
    \psi^{\text{CB/VB}}_{n,m,K,s}(\vec{r}) = \braket{\vec{r}|\psi^{\text{VB/CB}}_{n,m,s}(K)} = \alpha_{n,K,s}^{\text{VB/CB}} w_{n-1,m}(\vec{r}) u^{\text{A}}_{s} (\vec{r}) + \beta_{n,K,s}^{\text{VB/CB}} w_{n,m}(\vec{r}) u^{\text{B}}_{s} (\vec{r}).
\end{equation}
%%%%%%%%%%%%%%%%%%%%%%%%%%%%%%%%%%%%%%%%%%%%%%%%%%%%
Here, $w_{n,m}(\vec{r})=\braket{\vec{r}| n,m}$ is the real-space representation of the \ac{HO} orbital with quantum numbers $(n,m)$, and $u^{\text{A}}_{s} (\vec{r})$ and $u^{\text{B}}_{s} (\vec{r})$ are respectively the Bloch functions of the \ac{CB} and \ac{VB} subbands composing the basis of the Hamiltonians \eqref{eq:mDFmagneticH} with spin $s$. Similar expression is taken for \ac{LL} orbitals for the opposite valley. In the spirit of the envelope function approximation, the \ac{HO} orbitals give the long-range behavior of the wavefunction, and are treated as constant within the \ac{UC} of the lattice. On the other hand, the Bloch functions are lattice-periodic and describe the details of the wavefunction within each \ac{UC}. Since the magnetic length $l_B = \sqrt{\hbar/eB}$ significantly exceeds the lattice constant, the Coulomb interaction is considered in the long-distance limit and treated as effectively constant over the length scale of a single \ac{UC}.

Because each \ac{LL} orbital is a two-component wavefunction, the Coulomb matrix element $\braket{i,j|V_C|k,l}$ will consist of $16$ terms. One such term, written for the valley $K$, has the following form:
%%%%%%%%%%%%%%%%%%%%%%%%%%%%%%%%%%%%%%%%%%%%%%%%%%%%
\begin{eqnarray}
    V_{AAAA}^{i,j,k,l} &=& 
    \frac{e^2}{4\pi\varepsilon_0\varepsilon_r}
    (\alpha_i)^* (\alpha_j)^* \alpha_k \alpha_l
    \nonumber\\
    &\times& \sum_{\vec{R}_1} \sum_{\vec{R}_2}
    \frac{[w_{i}(\vec{R}_1)]^*
    [w_{j}(\vec{R}_2)]^*
    w_{k}(\vec{R}_2) w_{l}(\vec{R}_1)
    }{|\vec{R}_1-\vec{R}_2|} \times
    \int_{\Omega} { d\vec{r}_1 [u^A_{i}(\vec{r}_1)]^* u^A_{l}(\vec{r}_1)    }
    \int_{\Omega} { d\vec{r}_2 [u^A_{j}(\vec{r}_2)]^* u^A_{k}(\vec{r}_2)    },
    \nonumber
\end{eqnarray}
%%%%%%%%%%%%%%%%%%%%%%%%%%%%%%%%%%%%%%%%%%%%%%%%%%%%
The integral over all space has been explicitly replaced by a sum over unit cells, $\int_V d\vec{r} \rightarrow \sum_{\vec{R}}\int_{\Omega}d\vec{r}$, where $\vec{R}$ denotes the position of the unit cell center, and $\Omega$ represents the volume of the unit cell. For brevity, compact indices have been introduced, and the \ac{HO} orbitals are denoted by $w$, with the specific identification $i\equiv (n_i-1,m_i)$ applied to all four functions in this term. The corresponding term is labeled $AAAA$, indicating that the \ac{LL} orbital components originate from the subband $A$ in all four spinors. A total of 16 such terms arise, each assigned a distinct index sequence, such as $AAAB$, $ABBA$, etc. 

The decomposition of the Coulomb integral into long- and short-distance contributions results in the emergence of overlap integrals between the Bloch functions associated with orbitals $i$ and $l$, as well as $j$ and $k$. Due to the orthogonality of Bloch functions from different subbands, only 4 out of the 16 terms contribute---those corresponding to the index sequences $AAAA$, $ABBA$, $BAAB$, and $BBBB$. Additionally, spin selection rules impose that the two outer orbitals, $i$ and $l$, must possess the same spin, as must the two inner orbitals, $j$ and $k$.

With these selection rules in place, the Coulomb terms retain summations over the coordinates $\vec{R}_1$ and $\vec{R}_2$, which are subsequently evaluated by converting the discrete sums into integrals (omitting the Bloch functions at this stage). As a result, the full Coulomb matrix element is represented as a sum over four \ac{HO} Coulomb elements, $\braket{n_i,m_i;n_j,m_j|V_C|n_k,m_k;n_l,m_l}$, each corresponding to a different sequence of the indices $n$, and each weighted by an appropriate combination of the spinor coefficients $\alpha$ and $\beta$. The closed-form expressions for the \ac{HO} Coulomb elements can be found in Ref.~\cite{qdotsbook}.

%%%%%%%%%%%%%%%%%%%%%%%%%%%%%%%%%%%%%%
\section{2D electron gas in high magnetic field}  
\label{appendix:2DEG}
%%%%%%%%%%%%%%%%%%%%%%%%%%%%%%%%%%%%%%
%The emission spectrum of a neutral exciton (\ac{X0}) in the absence of a magnetic field has been extensively studied by many authors~\cite{CastroNeto2013, Thygesen2013, Louie2013, MacDonald2015, Louie2015, PingYuan2019, PingYuan2025}, including ourselves~\cite{bieniek2020band, bieniek2022nanomaterials}. We extend these zero-field studies to finite magnetic fields, building on our previous work on magnetoexcitons in MoS$_2$~\cite{Szulakowska2019magnetoX}. In particular, we incorporate Landau quantization in both conduction and valence bands and explore the effects of valley degrees of freedom, Coulomb interactions, and finite electron densities in the \ac{CB} on the excitonic spectrum. This framework allows us to investigate how excitonic states are modified by interactions with a gas of \acp{mDF} occupying \acp{LL}. The following subsections introduce the many-body Hamiltonian, outline the relevant excitonic configurations, and present the resulting optical spectra.

We now extend the zero-field studies of a neutral exciton (\ac{X0})~\cite{CastroNeto2013, Thygesen2013, Louie2013, MacDonald2015, Louie2015, PingYuan2019, PingYuan2025, bieniek2020band, bieniek2022nanomaterials} to finite magnetic fields, building on our previous work on magnetoexcitons in MoS$_2$~\cite{Szulakowska2019magnetoX}. The following subsections introduce the details on the many-body Hamiltonian, the relevant excitonic configurations, and the resulting optical spectra.

The single-particle \ac{LL} orbitals are populated with a total of $N = N_V + N_G$ electrons, where $N_V$ corresponds to the electrons filling the \ac{VB} \acp{LL}, and $N_G$ denotes the additional electrons brought into the \ac{CB} by the gate potential. The Hamiltonian of the interacting electrons can be written as
%%%%%%%%%%%% Eq  Hamiltonian  %%%%%%%%%%%%%
\begin{equation}
    \hat{H} = \sum_i E_i
    \hat{c}_i^\dagger\hat{c}_i
    + \frac{1}{2}\sum_{i,j,k,l} \braket{i,j|V_C|k,l} \hat{c}_i^\dagger\hat{c}_j^\dagger\hat{c}_k\hat{c}_l -\sum_{i,j} \braket{i|V_P|j}
    \hat{c}_i^\dagger\hat{c}_j.
\label{eq:InteractingHamiltonian}
\end{equation}
%%%%%%%%%%%%%%%%%%%%%%%%%%%%%
Here, the composite index $i = \{n, m, \tau, s, \text{CB/VB}\}$ encompasses the \ac{LL} orbital indices, valley, spin, and band index of the single-particle state (similarly for $j, k, l$). The second quantization operators $c_i$ ($c_i^\dagger$) anihilate (create) an electron in state $i$. $E_i$ denotes the single-particle energy of the \ac{LL} orbital $i$. The Coulomb matrix elements $\braket{i,j|V_C|k,l}$ describe the  interaction between electrons and are defined in the previous Section.  The scattering matrix elements $\braket{i|V_P|j}$ account for the interaction between electrons and the positive background, which ensures the overall charge neutrality of the system. For the $N_V$ electrons occupying the \ac{VB}, this background is provided by the lattice ions, while the charge of the remaining $N_G$ electrons is balanced by the positive gate potential. The resulting attractive potential can in principle be calculated by solving the Poisson equation, however we account for it in a model fashion as follows. First, all $N$ electrons are distributed  on the single-particle levels of both bands such that the resulting configuration has the lowest single-particle energy. The electrons are then replaced by positive charges, and this configuration is fixed. The matrix element of interaction between the actual electron and the positive charge distribution is therefore $\braket{i|V_P|j} = \sum_{m,\text{occ}} \braket{i,m|V_C|m,j}$, where the sum runs over all orbitals $m$ occupied by the positive charges. The electron-positive charge interactions include only the direct  terms, and interactions among the positive charges are neglected.

To explore the optical properties of the \ac{TMDC} as a function of $N$, we focus on interband magnetoexcitons, i.e., the excited states of the Hamiltonian~\eqref{eq:InteractingHamiltonian}, in which one electron is promoted from the \ac{VB} into the \ac{CB}, leaving behind a valence hole. We utilize the \ac{CI} approach, in which the Hamiltonian is written in a matrix form in the basis of configurations of all $N$ electrons distributed on a suitably large number $M$ of single-particle states. The computational cost of full \ac{CI} grows factorially with $N$ and $M$. To manage the computational complexity, the reduced basis for the low-energy excited states is built systematically in the language of electron-hole pair excitations~\cite{Hawrylak19972DEG, Yoo2023}. To this end, we choose the reference ground-state configuration $\ket{GS}$ as the configuration with the lowest total single-particle energy, and seek the low-energy eigenstates in the form
$\ket{\Psi_k} = \alpha^{(k)}\ket{GS} + \sum_{i,j}\beta_{i,j}^{(k)}\,\hat{c}_i^{\dagger}\hat{c}_j\ket{GS} +  \sum_{i,j,n,l}\gamma^{(k)}_{ijnl}\,\hat{c}_n^{\dagger}\hat{c}_l\hat{c}_i^{\dagger}\hat{c}_j\ket{GS}+\cdots$.
Here, $k$ enumerates the eigenstates of the $N$-electron Hamiltonian. The amplitudes $\alpha^{(k)}$, $\beta^{(k)}$, $\gamma^{(k)}$, as well as the energy $E_k$ of the state are established by diagonalizing the Hamiltonian matrix written in the basis of the pair excitation configurations consisting of up to $p_{max}$ pairs.

The parameter $p_{max}$ controls the convergence of the results in the desired energy range. In doped systems at zero magnetic field, such as the \ac{2DEG} in GaAs~\cite{Skolnick1987, Hawrylak1991FermiEdge, Brown1996, Brum1996} or in TMDCs~\cite{Sidler2017,FinleyDery2025}, the single-particle spectrum is typically not gapped. This leads to the appearance of signatures of strong correlations in the optical spectra of the exciton interacting with the \ac{2DEG} and requires $p_{max}$ to be large. On the other hand, the \ac{mDF} single-particle spectrum at high magnetic fields is characterized by a set of gaps, including the bandgap, the cyclotron gaps separating \acp{LL} with the same valley and spin degrees of freedom, and the spin-orbit, valley Zeeman, and spin Zeeman gaps related to the spin and valley physics. The existence of these gaps allows to classify the pair excitations into families with the same total single-particle energy, which naturally restricts the number $p_{max}$. If the gaps separating these families are larger than the Coulomb interactions connecting them, the Hamiltonian~(\ref{eq:InteractingHamiltonian}) can be diagonalized separately in each of the resulting relatively small Hilbert subspaces.

%%%%%%%%%%%%%%%%%%%%%%%%%%%%%%%%%%%%%%
\subsection{Details on the Magnetoexciton Spectra}  
\label{appendix:magnetoX*}
%%%%%%%%%%%%%%%%%%%%%%%%%%%%%%%%%%%%%%

The matrix elements of the many-body Hamiltonian connecting two electron-hole pair configurations, $\ket{i_1,j_i}$ and $\ket{i_2,j_2}$ are calculated as:
%%%%%%%%%%%%%%%%%%%%%%%%%%%%%
\begin{equation}
\braket{j_1,i_1|\hat{H}|i_2,j_2} = 
\delta_{i_1,i_2}\delta_{j_1,j_2}\left(E_{i_1}-E_{j_1}\right)
    -\braket{j_1,i_2|V_C|i_1,j_2}
        +\braket{j_1,i_2|V_C|j_2,i_1} 
    +\begin{bmatrix}
        \delta_{i_1,i_2}\sum_k^{N_{V}} \braket{k^{VB},j_1|V_C|k^{VB},j_2} \\ +\delta_{i_1,i_2}\sum_k^{N_{G}} \braket{k^{CB},j_1|V_C|k^{CB},j_2} \\
        -\delta_{j_1,j_2} \sum_k^{N_{V}} \braket{k^{VB},i_2|V_C|k^{VB},i_1} \\
        -\delta_{j_1,j_2} \sum_k^{N_{G}} \braket{k^{CB},i_2|V_C|k^{CB},i_1}
    \end{bmatrix}.
\label{eq:MagnetoX_BSEmatrix}
\end{equation}
%%%%%%%%%%%%%%%%%%%%%%%%%%%%%
The expectation value $\braket{GS|\hat{H}|GS}$ has been subtracted from the diagonal Hamiltonian elements. The first, diagonal term on the right-hand side of Eq.~(\ref{eq:MagnetoX_BSEmatrix}) accounts for the electron and hole single-particle energies (the energy of the hole enters with a minus sign as it is a missing electron). The following two terms account for the electron-hole direct and exchange interactions, respectively. Lastly, the terms in the square bracket account for the interactions of the pair with all other electrons, as well as with the positive background. The terms for the excited electron enter with negative signs, while the terms for the hole - with positive signs. Let us focus on the electron terms (the interpretation of the hole terms is analogous). The excited electron interacts with all the $N_V+N_G$ electrons in the system by direct and exchange terms, and with the same number of positive charges by direct terms only. In this model approach, the two types of direct terms cancel out, leaving only the exchange. The sums extend over the $N_V$ electrons in the \ac{VB} and the $N_G$ electrons in the $\nu=1$ configuration of the \ac{CB}, respectively, and we separated these two contributions for reasons explained below.

The diagonal matrix elements of the interacting Hamiltonian can be isolated, giving:
%%%%%%%%%%%%%%%%%%%%%%%%%%%%%
\begin{equation}
    \braket{j,i|\hat{H}|i,j} =
    (E_i + \Sigma_i )
    - (E_j + \Sigma_j) - \braket{j,i|V_C|i,j}
    +\braket{j,i|V_C|j,i}.
\label{eq:diagonal}
\end{equation}
%%%%%%%%%%%%%%%%%%%%%%%%%%%%%
Here, $\Sigma_i$ ($-\Sigma_j$) denotes the self-energy of the electron (hole) immersed in the electrons forming the $\ket{GS}$ in the presence of the positive background. As is evident from Eq.~(\ref{eq:MagnetoX_BSEmatrix}), $\Sigma_i = -\sum_k^{N_V+N_G} \braket{k,i|V_C|k,i} $ is built out of the exchange interactions with all electrons present in the system, and $\Sigma_j$ is built analogously. In principle, $\Sigma_i$ depends on the orbital index $m$ (included in the composite index $k$) within each \ac{LL}. However, in realistic systems the \ac{LL} degeneracy is of order $M\sim 10^{10}$, which results in a uniformity (translational invariance) of energies of orbitals within each \ac{LL}. With the chosen model degeneracies $M_{LL}=31$, this uniformity is not achieved resulting in pronounced edge effects. To avoid this, for each \ac{LL} the self-energy is evaluated only for the $m=0$ orbital and treat this value as a representative energy for the entire level.

%%%%%%%%%%%%% Fig %%%%%%%%%%%
\begin{figure}[tb]
\centering\includegraphics[width=0.55\linewidth]{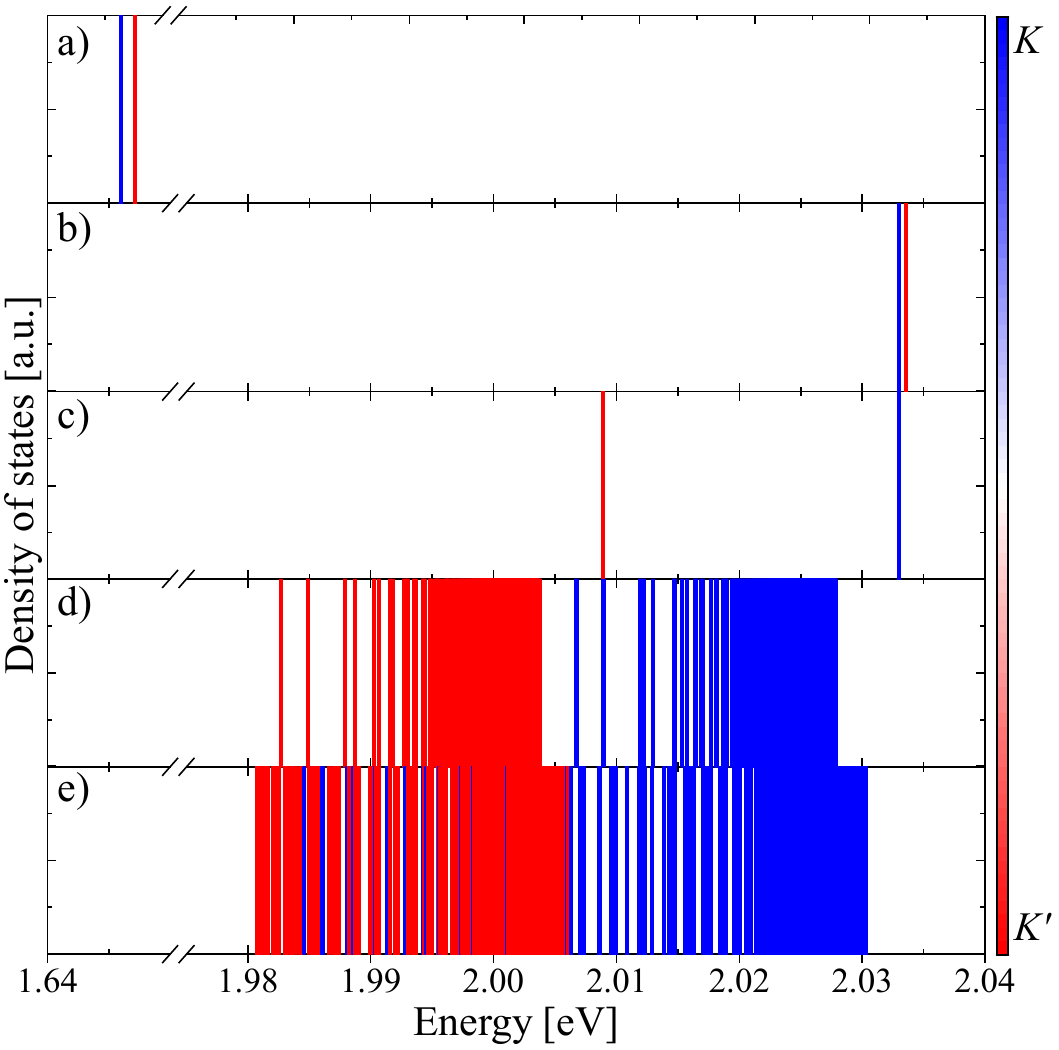}
    \caption{Evolution of the spectrum of interband exciton in the presence of a fully filled $n=0$ \ac{CB} ($\nu=1$), showing the progressive inclusion of Coulomb interactions. Color indicates valley. Panels (a–e) show excitonic states as vertical lines. (a) Noninteracting spectrum. (b) Inclusion of \ac{VB} self-energy corrections. (c) Inclusion of \ac{CB} self-energy effects from the filled $n=0$ \ac{LL}. (d) Inclusion of vertex corrections. (e) Full interacting spectrum with the correlation effects.}
\label{fig:X*spectra_evolution}
\end{figure}
%%%%%%%%%%%%%%%%%%%%%%%%%%%%%

Figure~\ref{fig:X*spectra_evolution} shows the evolution of the energy spectra of the interband exciton interacting with the $\nu=1$ configuration as successive components of the Coulomb interaction are included. Panel (a) presents the single-particle (non-interacting) pair energies. Pairs constructed in the $K$ valley (blue) have lower single-particle energy than those in the $K'$ valley (red), because the latter involve the electron occupying the $n=1$ \ac{LL} in $K'$, which lies higher in energy than the $n=1$ \ac{LL} in $K$ [see Fig.~\ref{fig:LLvsB}(b)]. This splitting arises from spin-orbit interactions and spin-valley Zeeman effects.

In panel (b), quasiparticle self-energy corrections due to interactions with the filled \ac{VB} (i.e., the $N_V$ electrons) are included. These corrections reduce the energy difference between the $K$ and $K'$ pairs and produce a significant blueshift of the spectrum. The blueshift is primarily due to the hole self-energy, which interacts more strongly with the filled \ac{VB} than the conduction electron. Although the self-energy term is negative, it enters the diagonal Hamiltonian with a negative sign (Eq.~\ref{eq:diagonal}), thereby increasing the energy of the pair.

Panel (c) introduces additional self-energy effects arising from the filled \ac{CB} ($\nu=1$ electrons). The energy of the $K'$ pairs decreases significantly, while the $K$-valley pair energies remain largely unaffected. This behavior is due to the exchange interaction of the excited electron with the $\nu=1$ state. If the electron resides in the $K'$ valley (together with the $\nu=1$ gas), the intravalley exchange correction is large (tens of meV) and negative, outweighing the single-particle energy cost (several meV). In contrast, an electron in the $K$ valley interacts with the $\nu=1$ electrons only via the much weaker intervalley exchange. Consequently, the ordering of the pair levels is inverted, with $K'$-valley configurations (red) forming the lower-energy manifold.

In panel (d), the attractive direct and repulsive exchange interactions between the electron and hole within each pair are included. These interactions produce a substantial redshift and lift the degeneracy among pairs, as electron-hole interactions are strongly configuration-dependent. Nevertheless, the low-energy portion of the spectrum remains dominated by $K'$-valley configurations.

Finally, panel (e) shows the spectra of correlated exciton states, $\ket{\Psi_k^{(ini)}}=\sum_{i,j} A_{i,j}^{(k)}\ket{i,j}$, obtained from exact diagonalization of the full Hamiltonian. Configuration mixing broadens the energy manifolds in each valley, resulting in overlap between the $K$ and $K'$ exciton families. Despite this, the ground state and several lowest-energy excited states remain primarily composed of $K'$-valley configurations.

%%%%%%%%%%%%%%%%%%%%%%%%%%%%%%%%%%%%%%
\subsection{Neutral Magnetoexciton Spectra}  
\label{appendix:neutralX}
%%%%%%%%%%%%%%%%%%%%%%%%%%%%%%%%%%%%%%

\begin{figure}[t]
\centering\includegraphics[width=0.5\linewidth]{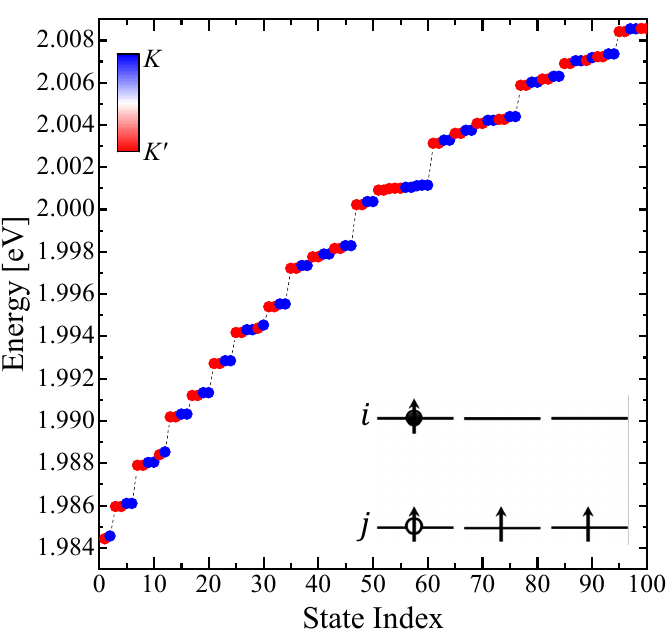}
    \caption
    {Excitonic spectra of the neutral exciton. The inset provides a schematic illustration of the corresponding optical complex.}
\label{fig:Xspectrum}
\end{figure}

For completeness we present a detailed study of a single neutral magnetoexciton, a linear combination of single electron–hole excitations (\ac{X0}). In this case, the number of excess conduction electrons is set to $N_G=0$, meaning the ground state $\ket{GS_0}$ of the system is formed by the Slater determinant of \ac{VB} fully occupied with electrons, while the \ac{CB} remains empty. 
In the magnetic field $B = 10$ T discussed throughout this analysis, single electron-hole pair excitations are then created in the form $|i,j\rangle = \hat{c}_i^{\dagger} \hat{c}_j |GS_0\rangle$. In the valley $K'$, the index $i$ labels the orbitals of the lowest $n=0$ \ac{LL} in the \ac{CB} with spin up. As illustrated in Fig.~\ref{fig:LLvsB}(b), this corresponds to the zeroth \ac{LL}, which forms a degenerate manifold of single-particle states with the lowest energy in the \ac{CB}. The index $j$ enumerates the orbitals of the $n=1$ \ac{LL} in the \ac{VB}, also with spin up. Conversely, in the valley $K$, the index $i$ labels the spin-down \ac{CB} orbitals belonging to the \ac{LL} with $n=1$, while the index $j$ traverses the spin-down zero-\ac{LL} orbitals in the \ac{VB}.  
One of the pair configurations is schematically shown in the inset of Fig.~\ref{fig:Xspectrum}, where the \ac{LL} orbitals for both the \ac{CB} electron and the hole in the \ac{VB} are labeled by the same number $m$. All possible pairs are generated, allowing both the \ac{CB} electron and the \ac{VB} hole to move independently along their respective \acp{LL} within each valley. With the \ac{LL} degeneracy set to $M_{LL}=31$ for both bands, an electron-hole pair basis of $960$ configurations per valley is obtained. Since in the model the Coulomb interactions do not change the overall valley quantum numbers of the pairs, the excitonic spectra are calculated separately for each valley. Pairs in which the electron and the hole occupy opposite valleys are excluded, as such configurations are optically inactive.

%%%%%%%%%%%%% Fig %%%%%%%%%%%
\begin{figure}[t]
\centering\includegraphics[width=0.55\linewidth]{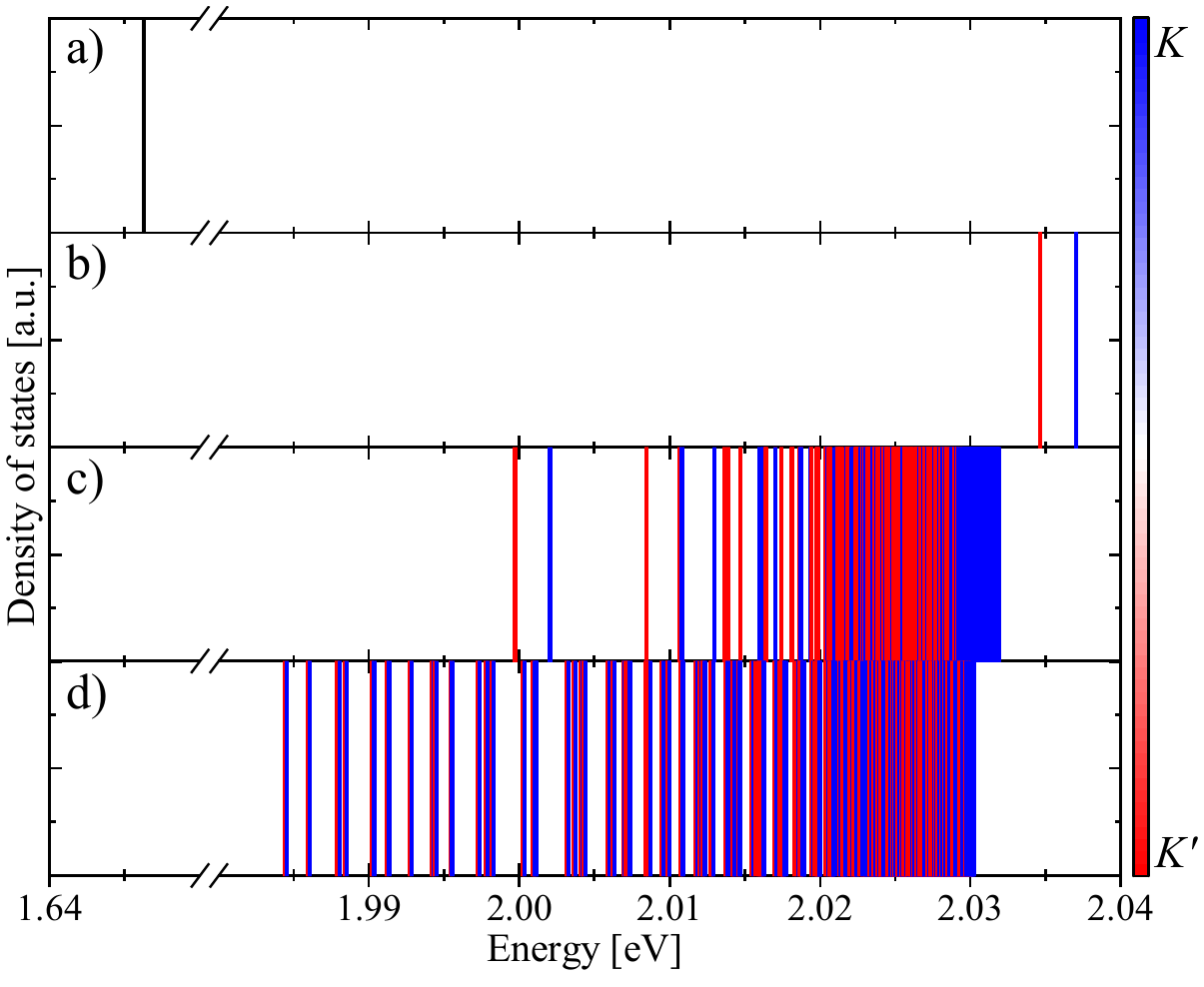}
    \caption
    {Evolution of the neutral exciton \ac{X0} spectrum in monolayer MoS$_2$ in a magnetic field, showing the sequential inclusion of Coulomb interactions within the \ac{mDF} model. Only spin-bright states are shown; color denotes valley. Vertical lines in panels (a–d) denote energies of excitonic states. (a) Non-interacting electron-hole pairs. (b) Inclusion of self-energy corrections. (c) Addition of vertex corrections. (d) Full interacting spectrum including correlation effects.}
\label{fig:Xspectra_evolution}
\end{figure}
%%%%%%%%%%%%%%%%%%%%%%%%%%%%%

The excitonic states are obtained by solving the \ac{BSE}, which, with the total energy of $\ket{GS_0}$ taken as reference, results in the set of equations:
%%%%%%%%%%%%%%%%%%%%%%%%%%%%%%%%%%%%%%
\begin{equation}
    \sum_{i',j'}  \Bigl[ ( E_{i'} + \Sigma_{i'})
    - (E_{j'} + \Sigma_{j'}) \Bigl] \delta_{i,i'} \delta_{j,j'} A^{(p)}_{i'j'} 
    + \sum_{i',j'} \Bigl( \braket{i,j'|V_C|i',j} 
    - \braket{i,j'|V_C|j,i'} \Bigl) A^{(p)}_{i'j'} = E_p A^{(p)}_{i,j}.
\end{equation}
%%%%%%%%%%%%%%%%%%%%%%%%%%%%%%%%%%%%%%
Here, $\Sigma_i$ is the self-energy of an electron/hole occupying the \ac{LL} orbital $i$/$j$ (discussed below), and $E_p$ and $A^{p}_{i,j}$ are, respectively, the eigenenergy and eigenvector components of the $p$-th excitonic state. In this equation, the Coulomb interactions enter in three distinct ways: through the self-energies, vertex corrections (diagonal matrix elements appearing for $i = i'$ and $j = j'$), and correlation effects, which lead to mixing between different pair configurations.

In analogy to the discussion presented previously,
we analyze the influence of each of these contributions on the excitonic spectrum, as visualized systematically in Fig.~\ref{fig:Xspectra_evolution}. Panel (a) shows the energies of electron-hole pair configurations in valley $K'$ without interaction effects, corresponding to differences between single-particle energies of the top valence $n=1$  and bottom conduction $n=0$ \acp{LL}, which are degenerate within each valley. The energy difference between these pairs from opposite valleys is negligibly small, hence these states are plotted in black. In panel (b), energy shifts due to the inclusion of self-energies $\Sigma_i$ are demonstrated. These self-energies account for interactions of an electron in orbital $i$ with all $N$ electrons and all positive charges forming the $\ket{GS_0}$ state. For the reasons explained in the previous Section, the self-energies account only for attractive exchange interactions with all electrons filling the \ac{VB}. For $i$ denoting a \ac{CB} orbital, the corresponding $\Sigma_i$ represents the total conduction-valence exchange energy, while for a \ac{VB} orbital $j$, $\Sigma_j$ is composed of exchange terms within the \ac{VB}. Consequently, $\Sigma_i$ is much smaller than $\Sigma_j$. Moreover, $\Sigma_i$ enters the \ac{BSE} with a positive sign, while $\Sigma_j$ enters with a negative sign, causing a blueshift of the pair energies. This pronounced blueshift, reflecting renormalization of quasiparticle energies due to interactions with the fully filled \ac{VB}, is clearly visible in panel (b). This step also lifts the degeneracy of the \acp{LL}, revealing a clearer energy separation between the $K$ and $K'$ valley configurations. Henceforth, configurations in different valleys are distinguished by color --- red for valley $K'$ and blue for valley $K$. In panel (c), the attractive electron-hole interaction as well as the relatively weak repulsive exchange interaction
within each electron-hole pair is accounted for.
While the exchange slightly shifts energies upward, the dominance of the direct attractive term causes a redshift of the spectrum. This stage further splits and reorganizes the excitonic states. Finally, panel (d) shows the energies of correlated \ac{X0} states incorporating all Coulomb scattering processes between different configurations. The resulting spectrum reflects the fully interacting spin-bright \ac{X0} response, with valley-specific character clearly visible through the color coding.

%%%%%%%%%%%%% Fig %%%%%%%%%%%
%%%%%%%%%%%%%%%%%%%%%%%%%%%%%

The final excitonic spectrum as a function of the \ac{X0} state index is shown in Fig.~\ref{fig:Xspectrum}. In contrast to extended systems at $B=0$~\cite{bieniek2020band, bieniek2022nanomaterials}, where low-energy states with different angular momentum are well-separated, the magnetoexciton spectrum is nearly linear, forming a broad band exceeding $20$~meV across the first $100$ states.

Furthermore, we evaluate the exciton binding energy, which quantifies the energy required to separate the electron and hole comprising the exciton. It is defined as the difference between the total energy of the non-interacting electron-hole pair (including self-energy corrections) and the \ac{GS} energy of the bound exciton:
%%%%%%%%%%%%%%%%%%%%%%%%%%%%%%%%%%%%%%%%%%%%%%%%%%%%
\begin{equation}
	E_b^{X^0} = \left[(E_e + \Sigma_e) - (E_h + \Sigma_h)\right] - E_0^{X^0},
\label{eq:MagnetoX_X0binding}
\end{equation}
%%%%%%%%%%%%%%%%%%%%%%%%%%%%%%%%%%%%%%%%%%%%%%%%%%%%
where $E_0^{X^0}$ is the \ac{GS} energy of the exciton. Using this definition, we obtain a binding energy of $E_b^{X^0} = 47$~meV, compared to $\sim 0.3$~eV in $B=0$~\cite{WuMacDonald2015}. This value reflects the stability of the exciton in the presence of both Coulomb interactions and an external magnetic field.

With this understanding of the excitonic spectra and binding energy, we next turn to computing the optical emission spectrum. The potential initial states in the recombination process are the eigenstates of our \ac{X0} in the form of linear combinations of electron-hole pairs, $\ket{\psi_p^{(ini)}}=\sum_{i,j}A^{(p)}_{i,j}\hat{c}^\dagger_{i}\hat{c}_{j}\ket{GS_0}$. 
The final state is simply the \ac{GS} of the $N$-electron system, $\ket{\psi^{(fin)}}=\ket{GS+0}$. In this case, therefore, the general Fermi's Golden Rule expression, can be rewritten as: 
%%%%%%%%%%%%%%%%%%%%%%%%%%%%%%%%%%%%%%%%%%%%%%%%%%%%%%%%%%%
\begin{equation}
	A\left(\hbar\omega\right) = \sum_k W_k \left| \left(\alpha^{CB}_{n_{CB}}\right)^*\beta^{VB}_{n_{VB}}\sum_{i}A^{(k)}_{i,i} \right|^2 \delta\left(E_k - \hbar\omega \right),
\label{eq:MagnetoX_X0emission}
\end{equation}
%%%%%%%%%%%%%%%%%%%%%%%%%%%%%%%%%%%%%%%%%%%%%%%%%%%%%%%%%%%
accounting for the fact that the \ac{GS} energy is taken as the reference. Here, $\alpha^{CB}_{n_{CB}}$ and $\beta^{VB}_{n_{VB}}$ denote the coefficients of the \ac{LL} spinor in the lowest \ac{CB} and \ac{VB} \ac{LL}, respectively, in each valley. The restriction to the coefficients $A^{(p)}_{i,i}$ under the summation follows from the optical selection rules, which permit contributions only from pair configurations with matching valley, spin, and orbital $m$ indices.  

%The calculated \ac{X0} emission spectrum is presented in Fig.~\ref{fig:Emission} for temperatures $T=1$~mK (a) and $T=5$~K (b). At the lower temperature, a single sharp emission line is observed, corresponding to the lowest-energy excitonic state. This state is composed of electron-hole pairs in the $K'$ valley and is predominantly formed by a pair configuration where the hole occupies the $(n=1, m=0)$ \ac{LL} and the electron resides in the $(n=0, m=0)$ state; i.e., the exciton is localized near the center of the system. As the effective temperature increases to $T=5$~K, a second emission line appears due to the thermal occupation of an almost-degenerate state. This higher-energy state is similar in its spectral content to the \ac{X0} \ac{GS}, but the pairs composing it reside in the valley $K$. According to the optical selection rules, photons emitted from the lowest-energy \ac{X0} state are $\sigma^+$ polarized, while those emitted from the higher-energy state are $\sigma^-$ polarized.

%%%%%%%%%%%%%%%%%%%%%%%%%%%%%%%%%%%%%%%%%%%%%%%%%%%%%

%%%%%%%%%%%%%%%%%%%%%%%%%%%%
\subsection{Negatively charged $K'$-valley magnetoexcitons}
\label{appendix:X-}
%%%%%%%%%%%%%%%%%%%%%%%%%%%%

We now provide a description of the calculations of the emission spectra of the negatively charged exciton, \ac{X-}. In this case, the single-particle states are populated with $N=N_V+1$ electrons. At the magnetic field $B=10$~T, the single electron in the \ac{CB} is placed on the lowest available \ac{LL} with $n=0$ in the valley $K'$. There exists, however, a manifold of $N$-electron configurations with equal single-particle energy, since the additional electron can be placed on any orbital $m$ of the zeroth \ac{LL}. As described earlier, a reference configuration can be taken in the form
%%%%%%%%%%%%%%%%%%%%%%%%%%%%%
\begin{equation}
    \ket{GS_1} = c^{\dagger}_{m=0}\ket{GS_0},
\end{equation}
%%%%%%%%%%%%%%%%%%%%%%%%%%%%%
where $\ket{GS_0}$ is the \ac{GS} configuration for the neutral exciton, i.e., one where the \ac{VB} is filled and the \ac{CB} is empty. The additional electron is placed on the orbital $m=0$ of the zeroth \ac{CB} \ac{LL}, and all configurations forming the degenerate manifold can be generated systematically by one-pair excitations within the zeroth \ac{LL}:
%%%%%%%%%%%%%%%%%%%%%%%%%%%%%
\begin{equation}
    \ket{m} = c^{\dagger}_m c_0 \ket{GS_1} \equiv c^{\dagger}_m \ket{GS_0}.
\end{equation}
%%%%%%%%%%%%%%%%%%%%%%%%%%%%%
Due to this ambiguity in the \ac{GS} of the $N$ electrons, no compensation of the additional \ac{CB} electron by a positive background is introduced.

In principle, one now has to consider how such configurations are mixed by the Coulomb interactions and/or positive background in the interacting Hamiltonian. It turns out, however, that in the zeroth-\ac{LL} approximation no off-diagonal Coulomb terms appear, and the final states are simply $\ket{\Psi_q^{(fin)}} = \ket{m}$. Their energies relative to that of $\ket{GS_0}$ are simply $E_q^{(fin)} = E_m + \Sigma_m$ (i.e., corrected by the self-energy $\Sigma_m$) and they are identical for all values of the index $m$. We find therefore, that the final states of the single quasielectron form a degenerate manifold with the degeneracy $M_{LL}$ equal to that of  the \ac{LL}.

\begin{figure}[t]
\centering\includegraphics[width=0.5\linewidth]{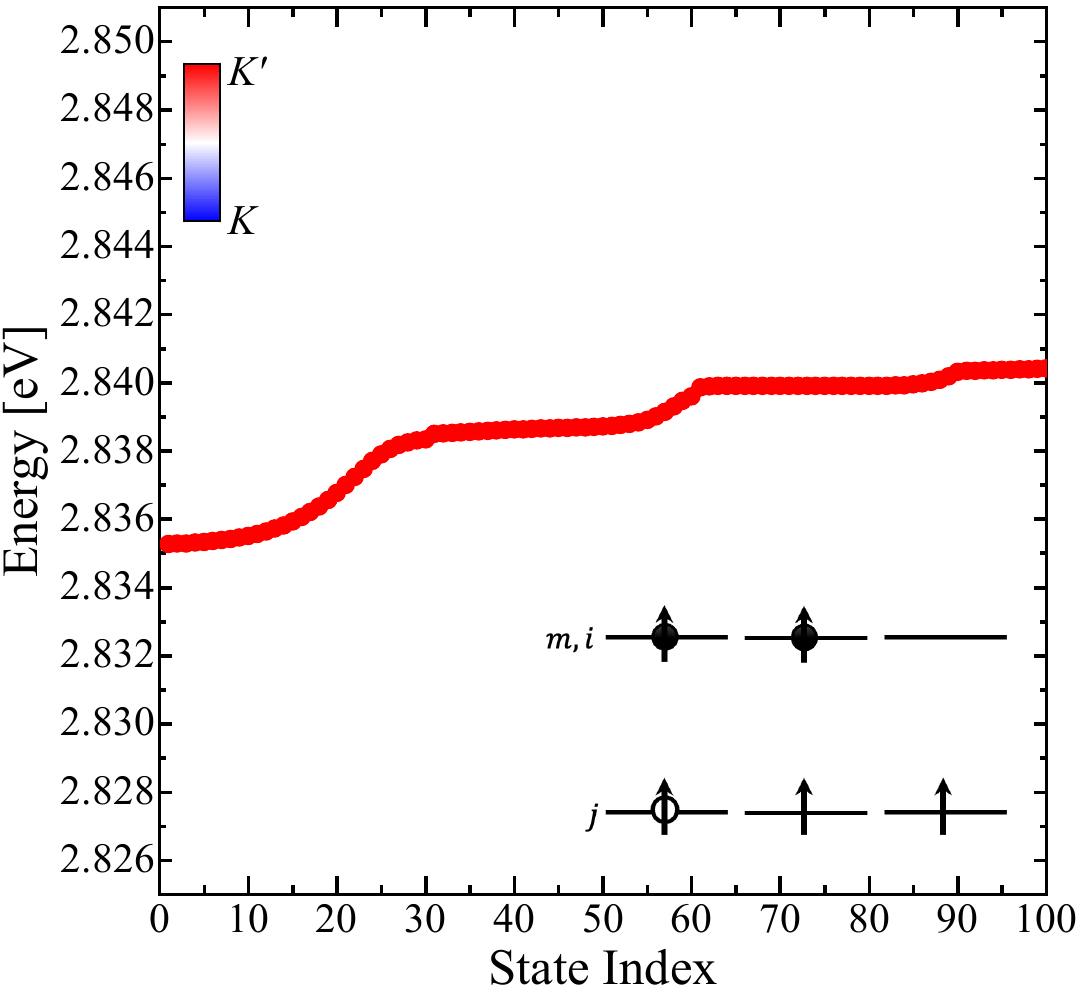}
    \caption
    {Energy spectra of the negatively charged exciton displayed within the same energy window as Fig.~\ref{fig:Xspectrum}. The inset provides a schematic illustration of the trion configuration in the valley $K'$.}
\label{fig:X-spectra}
\end{figure}

The initial-state $X^-$ configurations are created by adding an interband electron-hole pair excitation to the manifold $\ket{m}$. This treatment is restricted to single-valley trions, in which the pair excitation is placed in the valley $K'$, together with the resident electron. The configurations can be written as two-pair excitations
%%%%%%%%%%%%%%%%%%%%%%%%%%%%%
\begin{equation}
    \ket{m,i,j} = c^{\dagger}_i c_j c^{\dagger}_m c_0 \ket{GS_1}.
\end{equation}
%%%%%%%%%%%%%%%%%%%%%%%%%%%%%
Here, the first pair of creation and annihilation operators removes an electron from the orbital $j$ of the $n=1$ \ac{LL} in the \ac{VB} and replaces it on the orbital $i$ in the zeroth \ac{LL} in the \ac{CB}. The second creation-annihilation pair shifts the resident electron along the zeroth \ac{LL} in the \ac{CB}. To simplify further discussion, and to align with the choice of the positive background, the configurations are written in a simpler form of trions
%%%%%%%%%%%%%%%%%%%%%%%%%%%%
\begin{equation}
	\ket{m,i,j} = c^\dagger_{m} c^\dagger_{i} c_{j} \ket{GS_0}. 
\end{equation}
%%%%%%%%%%%%%%%%%%%%%%%%%%%%
relative to the unique ground state $\ket{GS_0}$. One representative configuration is shown schematically in the inset of Fig.~\ref{fig:X-spectra}. For a \ac{LL} degeneracy of $M$, a total of $\binom{M}{2} \times M$ trion configurations can be generated; for example, $M = 31$ gives $14415$ configurations. The \ac{X-} states are constructed as linear combinations of these configurations, $\ket{\Psi^{(ini)}_p} = \sum_{m,i,j} A^{(p)}_{m,i,j} \ket{m,i,j}$.

The many-body trion states are then obtained by solving the trion \ac{BSE}~\cite{Sadecka2024Trion} in the following form:
%%%%%%%%%%%%%%%%%%%%%%%%%%%%
\begin{equation}
\begin{split}
    &E_p^{(ini)} A^{(p)}_{m,i,j} = (E_m + E_i - E_j )  A^{(p)}_{m,i,j} \\
    &\quad+ 
    \sum_{m',i',j'}
    \left(
    \begin{array}{c}
\Sigma_{m',m}\delta_{i,i'}\delta_{j,j'} 
-\Sigma_{i',m}\delta_{i,m'}\delta_{j,j'}\\
    +\Sigma_{i',i} \delta_{m,m'}\delta_{j,j'}-\Sigma_{m',i}
    \delta_{m,i'}\delta_{j,j'}
    \\
-\Sigma_{j',j}\delta_{i,i'}\delta_{m,m'}    
    \end{array}
    \right)        
    A^{(p)}_{m',i',j'}
    + \sum_{m',i',j'} \left[
    \begin{array}{c}
    (V_{j,i',j',i} - V_{j,i',i,j}) \delta_{m,m'}\\
    +(V_{j,m',j',m} - V_{j,m',m,j'})\delta_{i,i'} \\
    -(V_{j,i',j',m} - V_{j,i',m,j'})\delta_{i,m'} \\
    -(V_{j,m',j',i} - V_{j,m',i,j'})\delta_{m,i'} \\
    -(V_{m',i',m,i} - V_{m',i',i,m})\delta_{j,j'} \\
    \end{array}
    \right] A^{(p)}_{m',i',j'},
\end{split}
\end{equation}
%%%%%%%%%%%%%%%%%%%%%%%%%%%%
where an abbreviated notation of Coulomb matrix elements has been used, $V_{j,i',j',i} \equiv \braket{j,i'|V_C|j',i}$ etc. 
Furthermore, we introduced generalized self-energy-like terms
$\Sigma_{p,q} = -\sum_m^{N_V+N_G} \braket{m,p|V_C|m,q}$
which include the possibility of single-particle scattering by the VB electrons.
In this equation, with the energy of $\ket{GS_0}$ taken as the reference, contributions from all three quasiparticles to the single-particle energies and self-energies, as well as the Coulomb interaction term, are included. The diagonal vertex correction, accounting for all interactions between quasiparticles in a given configuration $\ket{m,i,j}$, is obtained in the form:
%%%%%%%%%%%%%%%%%%%%%%%%%%%%
\begin{equation}
    E_V = 
    (V_{j,i,j,i} - V_{j,i,i,j}) 
    +(V_{j,m,j,m} - V_{j,m,m,j})
    -(V_{m,i,m,i} - V_{m,i,i,m}).
\end{equation}
%%%%%%%%%%%%%%%%%%%%%%%%%%%%
The first two groups of terms correspond, respectively, to exchange and direct electron-hole interactions, while the last term accounts for both exchange and direct electron-electron interactions. In our system, to a good approximation, the direct electron-electron repulsion nearly cancels with one of the direct electron-hole attraction terms. Furthermore, the electron-hole exchange terms are negligibly small. Therefore, the vertex correction is effectively composed of two significant contributions: one unit of electron-hole attraction and one unit of electron-electron exchange. The former enters identically into the energies of the neutral exciton, while the latter is unique to the \ac{X-} and results in a redshift of its emission energy, as discussed in the main text. 

The \ac{X-} energy spectrum $\{E_p^{(ini)}\}$, obtained by numerically diagonalizing the Hamiltonian matrix derived from the trion \ac{BSE}, is shown in Fig.~\ref{fig:X-spectra}. The energies of the first $100$ $X^-$ states are plotted as a function of their index, displaying a characteristic step-like structure. The chosen energy window matches that used to illustrate the \ac{X0} spectra in Fig.~\ref{fig:Xspectrum}. Evidently, addition of an extra electron into the system results in a flattening of the dispersion. This is due to the fact that the \ac{X-} states are constructed from many more low-energy configurations than the \ac{X0} states, since three quasiparticles (instead of two) can move along their respective \acp{LL}. This leads to additional correlation-induced renormalization of the trion dispersion. 

%%%%%%%%%%%%% Fig %%%%%%%%%%%
%%%%%%%%%%%%%%%%%%%%%%%%%%%%%

To evaluate the optical response, the emission spectrum associated with trion recombination is computed. A straightforward evaluation of the elements of the interband polarization then yields the following emission intensity as a function of photon energy:
%%%%%%%%%%%%%%%%%%%%%%%%%%%%
\begin{equation}
    A\left(\hbar\omega\right) = 
    \sum_{k,q} W_k \delta\left(E_k^{(ini)} - E_q^{(fin)} - \hbar\omega \right)
    \left| \left(\alpha^{CB}_{n=0}\right)^*\beta^{VB}_{n'=1} \right|^2 \times \left| \sum_{m,i}A^{(k)}_{m,i,m}-\sum_{m,i}A^{(k)}_{m,i,i} \right|^2,
\label{eq:MagnetoX_X-emission}
\end{equation}
%%%%%%%%%%%%%%%%%%%%%%%%%%%%
where $W_k$ is the thermal occupation factor of the initial trion state $\ket{\Psi^{(ini)}_k}$.

Three fundamental properties of the \ac{X-} emission spectrum spectrum can be observed:
(i) a pure $\sigma^+$ polarization of the emission,
(ii) a redshift of the main emission line relative to that of \ac{X0}, but not as large as that characterizing the magnetoexciton interacting with the $\nu=1$ configuration, and
(iii) a broadening of the spectrum at $T=5$~K, but not at $T=1$~mK. 

The polarized character of the \ac{X-} emission is a simple consequence of the three quasiparticles residing in the $K'$ valley. As discussed above, the redshift relative to the \ac{X0} line originates from more complex vertex corrections and correlation effects compared to those present in \ac{X0}, and in particular, the attractive  electron-electron exchange term characterizing the \ac{X-} initial state energies. Finally, the broadening of the \ac{X-} spectrum at higher temperature, seen in contrast to the \ac{X0} emission, can be understood by comparing the exciton spectra in Fig.~\ref{fig:Xspectrum} to the \ac{X-} spectra in Fig.~\ref{fig:X-spectra}. We find that the \ac{X-} spectra are composed of many low-lying states, as opposed to the \ac{X0} case, where the dispersion is much steeper. While the temperature $T=1$ mK is not enough to populate any \ac{X-} states beyond its \ac{GS}, at $T=5$~K, the low-lying excited states of this complex become thermally populated and result in the observed broadening of the emission spectrum.

%\newpage
\bibliography{Bibliography}